\newif\iffront\fronttrue     

\newif\ifpdone\pdonetrue     

\newif\ifmy\mytrue     
\newif\ifmy\myfalse   
\documentclass{article}
\usepackage{fullpage,amsmath,multicol,graphicx,verbatim,color}

\ifmy\usepackage{myhead,epsfig}
\else\usepackage{epsfig}\fi

\ifmy
\else\fi

\def\proof{\noindent{\bf Proof.}}

\def\OUT#1{\mbox{${\tt Output}(#1)$}}

\long\def\comment#1{~\\{}\fbox{{\bf comment}\footnotemark}%
 \footnotetext{\underline{to be dwelt upon:}\\{}#1}\\{}}
\long\def\comment#1{}

\long\def\done#1#2{\if#11{\ifpdone\footnote{{\huge\bf done}\ {#2}}\fi}
            \else\ifpdone\footnote{{\huge\bf do-it}\ {#2}}\fi\fi}

\def\mathbf#1{\mbox{$\boldmath #1$}}
\def\mbf#1{\mbox{$\boldmath \mbox{#1}$}}
\def\une{\mbox{${\mathchoice{\rm 1\mskip-4mu l}{\rm 1\mskip-4mu l}%
                         {\rm 1\mskip-4.5mu l}{\rm 1\mskip-5mu l}}$}}

\newcount \WIDTH  
\newcount \HEIGHT 
\newcount \Xcur   
\newcount \Ycur   
\newcount \HALF   
\newcount \DBL    
\newcount \QUA    
\newcount \TreeH
\newcount \TreeW

\long\def\commentt#1{}

\commentt{%
\def\treeXZ#1#2#3#4#5#6#7#8#9%
{\setlength{\unitlength}{1pt}\HEIGHT=#1\multiply\HEIGHT by 3
 \DBL=#1\multiply\DBL by 3\divide\DBL by 2%
\advance\HEIGHT by\DBL\advance\HEIGHT by 4
\begin{picture}(0,\HEIGHT)\put(0,\HEIGHT){%
\begin{picture}(0,0)\thinlines
 \Ycur=0
 \put(0,-\Ycur){\makebox(0,0)[c]{$\bullet$}}
 \put(0,-\Ycur){\makebox(0,0)[lb]{#2}}%
\advance \Ycur by #1
\end{picture}}\end{picture}}

{\allowdisplaybreaks  
\begin{align*}
    xxx & \Leftrightarrow
\\  &
\end{align*}
}
} 

\newcommand{\vPICTURE}[5]%
{\setlength{\unitlength}{1pt}\HEIGHT=#5\multiply\HEIGHT by #2%
 \WIDTH=14\multiply\WIDTH by #3\advance\HEIGHT by \WIDTH%
 \WIDTH=#5\multiply\WIDTH by #4%
\begin{picture}(\WIDTH,\HEIGHT)
  \divide\WIDTH by 2%
  \put(\WIDTH,\HEIGHT){#1{#5}}
\end{picture}}

\def\toLeft{r}
\def\toRight{l}

\def\Varrow#1#2{{\hbox{\(\left#1\vbox to #2{}\right.%
               \nulldelimiterspace=0pt\mathsurround=0pt\)}}}

\def\SouthEast#1#2#3
{\begin{picture}(0,0)\thicklines
 \put(0,0){\vector(1,-1){#1}}
 \HALF=#1 \divide\HALF by 2
 \put(\HALF,-\HALF){\makebox(0,0)[b#3]{\underline{#2}}}
\end{picture}}

\def\SouthWest#1#2#3
{\begin{picture}(0,0)\thicklines
 \put(0,0){\vector(-1,-1){#1}}
 \HALF=#1 \divide\HALF by 2
 \put(-\HALF,-\HALF){\makebox(0,0)[b#3]{\underline{#2}}}
\end{picture}}

\def\SouthEEast#1#2#3
{\begin{picture}(0,0)\thicklines
\QUA=#1 \multiply \QUA by 4
 \put(0,0){\vector(4,-1){\QUA}}
 \HALF=#1 \divide\HALF by 2\QUA=#1 \multiply \QUA by 2
 \put(\QUA,-\HALF){\makebox(0,0)[b#3]{{#2}}}
\end{picture}}

\def\SouthWWest#1#2#3
{\begin{picture}(0,0)\thicklines
\QUA=#1 \multiply \QUA by 4
 \put(0,0){\vector(-4,-1){\QUA}}
 \HALF=#1 \divide\HALF by 2\QUA=#1 \multiply \QUA by 2
 \put(-\QUA,-\HALF){\makebox(0,0)[b#3]{{#2}}}
\end{picture}}

\newcommand{\hitEast}
{\begin{picture}(0,0)\thicklines%
 \put(-8,4){\vector(2,-1){8}}
\end{picture}}
\newcommand{\hitWest}
{\begin{picture}(0,0)\thicklines%
 \put(8,-4){\vector(-2,1){8}}
\end{picture}}

\newcommand{\dashV}[2]
{\begin{picture}(0,0)\thinlines%
\TreeW=#1  \multiply \TreeW by 2
\TreeH=#2 \divide \TreeH by \TreeW
 \multiput(0,0)(0,\TreeW){\TreeH}{\line(0,1){#1}}
\end{picture}}

\newcommand{\dEast}[1]
{\begin{picture}(#1,0)\thicklines%
 \HALF=#1\divide\HALF by 2\DBL=\HALF\divide\DBL by 2%
 $\bezier{\DBL}(0,0)(\HALF,\DBL)(#1,0)$%
 \put(#1,0){\hitEast}
\end{picture}}

\newcommand{\Darrow}[1]
{\begin{picture}(0,0)\thicklines%
 \HALF=#1\divide\HALF by 2%
 $\bezier{\HALF}(-#1,0)(0,\HALF)(#1,0)$%
 \put(#1,0){\hitEast}
\end{picture}}

\newcommand{\Dbarrow}[1]
{\begin{picture}(0,0)\thicklines%
 \HALF=#1\divide\HALF by 2%
 $\bezier{\HALF}(-#1,0)(0,-\HALF)(#1,0)$%
 \put(-#1,0){\hitWest}
\end{picture}}

\def\CircNode#1
{\circle{20}
 \makebox(0,0)[c]{#1}}

\def\CCircNode#1
{\circle{20}\circle{16}
 \makebox(0,0)[c]{#1}}

\def\South#1#2#3
{\begin{picture}(0,0)\thicklines%
 \put(0,0){\vector(0,-1){#1}}
 \HALF=#1 \divide\HALF by 2
 \put(0,-\HALF){\makebox(0,0)[b#3]{\underline{#2}}}
\end{picture}}

\def\cSouth#1#2#3
{\begin{picture}(0,0)\thicklines%
 \HALF=#1 \advance \HALF by -20 
 \put(0,10){\line(0,1){\HALF}}
 \HALF=#1 \divide\HALF by 2
 \put(0,\HALF){\makebox(0,0)[b#3]{\underline{#2}}}
 \put(0,24){\vector(0,-1){14}}
\end{picture}}

\newcommand{\SouthB}[3]
{\begin{picture}(0,0)\thicklines%
 \put(0,0){\line(0,-1){#1}}
 \HALF=#1 \divide\HALF by 2
 \put(0,-\HALF){\makebox(0,0)[b#3]{\underline{#2}}}
\end{picture}}

\newcommand{\cSouthWest}[3]
{\begin{picture}(0,0)\thicklines
 \HALF=#1 \advance \HALF by -16 
 \put(9,9){\line(1,1){\HALF}}
 \HALF=#1 \divide\HALF by 2
 \put(\HALF,\HALF){\makebox(0,0)[b#3]{\underline{#2}}}
 \put(22,22){\vector(-1,-1){14}}
\end{picture}}

\newcommand{\cNorthEast}[3]
{\begin{picture}(0,0)\thicklines%
 \HALF=#1 \advance \HALF by -15 
 \put(8,8){\line(1,1){\HALF}}
 \HALF=#1 \divide\HALF by 2
 \put(\HALF,\HALF){\makebox(0,0)[b#3]{\underline{#2}}}
 \HALF=#1 \advance \HALF by -22 
 \put(\HALF,\HALF){\vector(1,1){14}}
\end{picture}}

\newcommand{\cSouthEast}[3]
{\begin{picture}(0,0)\thicklines%
 \HALF=#1 \advance \HALF by -16 
 \put(-9,9){\line(-1,1){\HALF}}
 \HALF=#1 \divide\HALF by 2
 \put(-\HALF,\HALF){\makebox(0,0)[b#3]{\underline{#2}}}
 \put(-22,22){\vector(1,-1){14}}
\end{picture}}

\def\ccSouthWest#1#2#3
{\begin{picture}(0,0)\thicklines%
 \DBL=#1 \multiply\DBL by 2 \advance\DBL by -27
 \put(18,9){\line(2,1){\DBL}}
 \DBL=#1 \advance\DBL by -7
 \HALF=#1 \divide\HALF by 2
 \put(\DBL,\HALF){\makebox(0,0)[b#3]{\underline{#2}}}
 \put(22,11){\vector(-2,-1){13}}
\end{picture}}

\def\ccSouthEast#1#2#3
{\begin{picture}(0,0)\thicklines%
 \DBL=#1 \multiply\DBL by 2 \advance\DBL by -27
 \put(-18,9){\line(-2,1){\DBL}}
 \DBL=#1 \advance\DBL by -7
 \HALF=#1 \divide\HALF by 2
 \put(-\DBL,\HALF){\makebox(0,0)[b#3]{\underline{#2}}}
 \put(-22,11){\vector(2,-1){13}}
\end{picture}}

\newcommand{\WEarrow}[3]
{\begin{picture}(0,0)\thicklines%
 \QUA=#2\divide\QUA by 2%
 \DBL=#1\advance\DBL by -9%
 \HALF=#1\advance\HALF by -#2\advance\HALF by 11
 $\bezier{200}(-\DBL,5)(-\HALF,\QUA)(0,\QUA)$%
  \advance\DBL by -6\advance\HALF by -7
 $\bezier{200}(0,\QUA)(\HALF,\QUA)(\DBL,8)$%
 \put(\DBL,8){\vector(2,-1){6}}
 \put(0,\QUA){\makebox(0,0)[b]{\raisebox{3pt}{#3}}}
\end{picture}}

\newcommand{\EWarrow}[3]
{\begin{picture}(0,0)\thicklines%
 \QUA=#2\divide\QUA by 2%
 \DBL=#1\advance\DBL by -9%
 \HALF=#1\advance\HALF by -#2\advance\HALF by 9
 $\bezier{200}(\DBL,-5)(\HALF,-\QUA)(0,-\QUA)$%
  \advance\DBL by -6\advance\HALF by -5
 $\bezier{200}(0,-\QUA)(-\HALF,-\QUA)(-\DBL,-8)$%
 \put(-\DBL,-8){\vector(-2,1){6}}
 \put(0,-\QUA){\makebox(0,0)[t]{\raisebox{-6pt}{#3}}}
\end{picture}}

\newcommand{\WEloop}[2]
{\begin{picture}(0,0)\thicklines%
 \HALF=#1\divide\HALF by 2%
 $\bezier{200}(-5,9)(-\HALF,#1)(0,#1)$%
 \put(9,17){\vector(-1,-2){4}}
 $\bezier{200}(0,#1)(\HALF,#1)(8,15)$%
 \put(0,#1){\makebox(0,0)[b]{\raisebox{2pt}{#2}}}
\end{picture}}

\newcommand{\dotSE}[1]
{\begin{picture}(0,0)\thicklines%
 \HALF=#1\divide\HALF by 2\DBL=\HALF\divide\DBL by 2%
 $\bezier{\DBL}(0,0)(\HALF,-\HALF)(#1,-#1)$%
 \DBL=#1\advance\DBL by -8%
 \put(\DBL,-\DBL){\vector(1,-1){8}}
\end{picture}}

\def\cdotSW#1#2#3
{\begin{picture}(0,0)\thicklines%
 \DBL=#1\divide\DBL by 4%
 \HALF=#1\advance\HALF by -8%
 $\bezier{\DBL}(10,10)(16,16)(\HALF,\HALF)$%
 \HALF=#1\divide\HALF by 2%
 \put(\HALF,\HALF){\makebox(0,0)[#3]{#2}}%
 \put(16,16){\vector(-1,-1){8}}%
\end{picture}}

\def\cdotSE#1#2#3
{\begin{picture}(0,0)\thicklines%
 \DBL=#1\divide\DBL by 4%
 \HALF=#1\advance\HALF by -8%
 $\bezier{\DBL}(-10,10)(-16,16)(-\HALF,\HALF)$%
 \HALF=#1\divide\HALF by 2%
 \put(-\HALF,\HALF){\makebox(0,0)[#3]{#2}}%
 \put(-16,16){\vector(1,-1){8}}%
\end{picture}}                             

\def\cdotWSE#1
{\begin{picture}(0,0)\thicklines%
 \DBL=#1\divide\DBL by 2%
 \HALF=#1\divide\HALF by 2%
 $\bezier{\DBL}(-1,1)(-\HALF,\HALF)(-#1,#1)$%
 $\bezier{\DBL}(1,1)(\HALF,\HALF)(#1,#1)$%
 \put(-8,8){\vector(1,-1){8}}%
 \put(8,8){\vector(-1,-1){8}}%
\end{picture}}

\newcommand{\cNorthWest}[3]
{\begin{picture}(0,0)\thicklines%
 \HALF=#1 \advance \HALF by -15 
 \put(-8,8){\line(-1,1){\HALF}}
 \HALF=#1 \divide\HALF by 2
 \put(-\HALF,\HALF){\makebox(0,0)[b#3]{\underline{#2}}}
 \HALF=#1 \advance \HALF by -22 
 \put(-\HALF,\HALF){\vector(-1,1){14}}
\end{picture}}

\newcommand{\SouthTwo}[3]
{\begin{picture}(0,0)\thicklines%
 \multiput(0,0)(#1,0){2}{\vector(0,-1){#1}}
 \HALF=#1 \divide \HALF by 2
 \put(  0,-\HALF){\makebox(0,0)[bl]{\underline{#2}}}
 \put(\HALF,-4){\makebox(0,0)[c]{\BF{\ \ldots}}}
 \put( #1,-\HALF){\makebox(0,0)[bl]{\underline{#3}}}
\end{picture}}

\newcommand{\HLine}[2]
{\begin{picture}(0,0)\thicklines%
 \put(0,0){\circle{16}}
 \put(0,0){\makebox(0,0)[c]{{\tiny #2}}}
 \HALF=#1 \advance \HALF by -8
 \put(-#1,0){\vector( 1,0){\HALF}}
 \put( #1,0){\vector(-1,0){\HALF}}
\end{picture}}

\newcommand{\DashBox}[6]
{\begin{picture}(0,0)\thicklines%
 \put(-#1,0){\South{#1}{#4}{\toLeft}}
 \put(  0,0){\SouthTwo{#1}{#5}{#6}}
 \HALF=#1 \divide \HALF by 3 \advance \HALF by #1
 \DBL=\HALF \multiply \DBL by 2
 \put(-\HALF,0){\dashbox{2}(\DBL,#2)[c]{#3}}
\end{picture}}

\newcommand{\DashBoxB}[6]
{\begin{picture}(0,0)\thicklines%
 \put(-#1,0){\SouthB{#1}{#4}{\toLeft}}
 \put(  0,0){\SouthTwo{#1}{#5}{#6}}
 \HALF=#1 \divide \HALF by 3 \advance \HALF by #1
 \DBL=\HALF \multiply \DBL by 2
 \put(-\HALF,0){\dashbox{2}(\DBL,#2)[c]{#3}}
\end{picture}}

\newcommand{\DashBoxX}[6]
{\begin{picture}(0,0)\thicklines%
 \put(-#1,0){\South{#1}{#4}{\toLeft}}
 \put(  0,0){\SouthTwo{#1}{#5}{#6}}
 \HALF=#1 \divide \HALF by 3 \advance \HALF by #1
 \DBL=\HALF \multiply \DBL by 2 \advance\DBL by #1
 \put(-\HALF,0){\dashbox{2}(\DBL,#2)[c]{#3}}
\end{picture}}

\newcommand{\NablaBox}[9]
{\begin{picture}(0,0)\thicklines%
 \put(-#1,-1){\makebox(0,0)[c]{$\bullet$}} 
 \put(-#1,0){\SouthB{#1}{#7}{\toLeft}}
 \put(  0,0){\SouthTwo{#1}{#8}{#9}}
 \HALF=#1 \advance \HALF by 12
 \put(0,\HALF){\DashBox{#1}{16}{#3}{#4}{#5}{#6}}
 \HALF=#1 \advance \HALF by 42 \DBL=#1 \multiply \DBL by 2
 \put(-\DBL,0){\multiply\DBL by 2\framebox(\DBL,\HALF){}}
 \advance \HALF by 3
 \put(0,\HALF){\makebox(0,0)[b]{#2}}
\end{picture}}

\newcommand{\FlatBoxB}[5]
{\begin{picture}(0,0)\thicklines%
 \put(-#1,0){\circle{16}}
 \put(-#1,0){\makebox(0,0)[c]{!}}
 \put(-#1,8){\makebox(0,0)[bl]{#2}}
 \HALF=#1 \advance \HALF by -8
 \put(0,0){\line(-1,0){\HALF}}
 \put(0,0){\line(1,0){#1}}
 \put(-#1,-8){\SouthB{\HALF}{#3}{\toLeft}}
 \put(0,0){\SouthTwo{#1}{#4}{#5}}
\end{picture}}

\newcommand{\FlatBoxBB}[2]
{\begin{picture}(0,0)\thicklines%
 \put(-#1,0){\circle{16}}
 \put(-#1,0){\makebox(0,0)[c]{!}}
 \put(-#1,8){\makebox(0,0)[bl]{#2}}
 \HALF=#1 \advance \HALF by -8
 \put(0,0){\line(-1,0){\HALF}}
 \put(0,0){\line(1,0){#1}}
 \put(-#1,-8){\line(0,-1){\HALF}}
 \multiput(0,0)(#1,0){2}{\line(0,-1){#1}}
 \HALF=#1 \divide \HALF by 2
 \put(\HALF,-4){\makebox(0,0)[c]{\BF{\ \ldots}}}
\end{picture}}

\newcommand{\seq}[2]{\mbox{$#1\,\vdash #2$}}     

\def\QED{\hspace*{\fill}\vrule height6pt width6pt depth0pt}

\def\clap#1{\hbox to 0pt{\hss#1\hss}}

\newcommand{\son}[2]
{\mbox{${\mbox{#1}}\over{\mbox{#2}}$}}

\newcommand{\pushb}[1]
{\mbox{${}\atop{\mbox{#1}}$}}

\def\sonN#1#2#3
{\thicklines%
\begin{tabular}[b]{@{}c@{}l@{}}#2 &\\ \multispan1\hrulefill&\omit\\%
  #3 &\rlap{\(\smash{\mathop{\phantom{o}}\limits^{\,\mbox{#1}}}\)}%
 \end{tabular}}

\newdimen \LtoR

\def\sonNl#1#2#3
{\settowidth\LtoR{#3}\LtoR=0.5\LtoR\hspace*{\LtoR}%
 \clap{\sonN{#1}{#2}{#3}}}

\def\sonNr#1#2#3
{\clap{\sonN{#1}{#2}{#3}}%
 \settowidth\LtoR{#3}\LtoR=0.5\LtoR\hspace*{\LtoR}}

\def\sonNc#1#2#3
{\clap{\sonN{#1}{#2}{#3}}}

\def\sonNcc#1#2#3
{\settowidth\LtoR{#3}\LtoR=0.5\LtoR\hspace*{\LtoR}%
 \clap{\sonN{#1}{#2}{#3}}%
 \settowidth\LtoR{#3}\LtoR=0.5\LtoR\hspace*{\LtoR}}

\newcommand{\Cson}[4]
{\begin{picture}(0,0)\thinlines
 \HALF=#2 \divide\HALF by 2
 \settowidth\LtoR{#3} \LtoR=0.5\LtoR \advance\LtoR #1pt
 \put( 0,-\HALF){\makebox(0,0)[cr]{\vbox{\hrule width\LtoR}}}
 \put( 0,-\HALF){\makebox(0,0)[cl]{\vbox{\hrule width\LtoR}}}
 \put( 0,-\HALF){\makebox(0,0)[cl]{\hspace*{\LtoR}{#4}}}
 \put(0,0){\makebox(0,0)[c]{#3}}
\end{picture}}
\newcommand{\Sons}[5]
{\begin{picture}(0,0)
 \HALF=#2 \divide\HALF by 2
 \settowidth\LtoR{#3} \LtoR=0.5\LtoR \advance\LtoR #1pt
 \put( 0,-\HALF){\makebox(0,0)[cr]{\vbox{\hrule width\LtoR}}}
 \settowidth\LtoR{#4} \LtoR=0.5\LtoR \advance\LtoR #1pt
 \put( 0,-\HALF){\makebox(0,0)[cl]{\vbox{\hrule width\LtoR}}}
 \put( 0,-\HALF){\makebox(0,0)[cl]{\hspace*{\LtoR}{#5}}}
 \put(-#1,0){\makebox(0,0)[c]{#3}}
 \put( #1,0){\makebox(0,0)[c]{#4}}
\end{picture}}

\def\SonTri#1#2#3#4#5#6
{\begin{picture}(0,0)
 \HALF=#2 \divide\HALF by 2
 \settowidth\LtoR{#3} \LtoR=0.5\LtoR \advance\LtoR #1pt
 \put( 0,-\HALF){\makebox(0,0)[cr]{\vbox{\hrule width\LtoR}}}
 \settowidth\LtoR{#5} \LtoR=0.5\LtoR \advance\LtoR #1pt
 \put( 0,-\HALF){\makebox(0,0)[cl]{\vbox{\hrule width\LtoR}}}
 \put( 0,-\HALF){\makebox(0,0)[cl]{\hspace*{\LtoR}{#6}}}
 \put(-#1,0){\makebox(0,0)[c]{#3}}
 \put(  0,0){\makebox(0,0)[c]{#4}}
 \put( #1,0){\makebox(0,0)[c]{#5}}
\end{picture}}

\def\gsep{\raise-2pt\hbox{\vrule\vbox{\hrule\kern1em%
                          \hbox to 3.5pt{}\hrule}\vrule}}

\def\llto{\mathbin{-\mkern-3mu\circ}}            
\newcommand{\limply}[2]{\mbox{$(#1 \llto #2)$}}  

\newcommand{\bang}[1]{\mbox{{\rm\,!}$#1$}}       

\newcommand{\lvariant}[3]{\mbox{$(#1 \llto (#2 \oplus #3))$}}

\newcount\PLv\newcount\PLw\newcount\PLx\newcount\PLy
\newdimen\PLyy\newdimen\PLX \newbox\PLdot \setbox\PLdot\hbox{\tiny.}
\def\scl{.08} 
\def\PLot#1{\PLx`#1\advance\PLx-42\PLy\PLx\PLv\PLx\divide\PLy9%
 \PLw\PLy\multiply\PLw9\advance\PLx-\PLw\advance\PLx-4\PLy-\PLy%
 \advance\PLy4\PLX=\the\PLx pt\advance\PLyy\the\PLy pt\wd%
 \PLdot=\scl\PLX\raise\scl\PLyy\copy\PLdot}
\def\draw#1{\ifx#1\end\let\next=\relax\else\PLot#1%
            \let\next=\draw\fi\next}
\def\IA{\hbox{\PLyy=70pt\draw :::;DMV_gqppyyyyyooooxxxnnwvlutkja%
  WNE=5-./99:::CCCC:::99/..--544=EENWWaajjjkktttttttNNNVVVVVVVV\end
  \hskip7pt}} 
\newbox\IAbox\setbox\IAbox\IA

\newtheorem{theorem}{Theorem}[section]
\newtheorem{lemma}{Lemma}[section]
\newtheorem{corollary}{Corollary}[section]
\newtheorem{proposition}{Proposition}[section]
\newtheorem{Defi}{Definition}[section]
\newenvironment{definition}{\begin{Defi}\em }{\end{Defi}}
\newtheorem{Exa}{Example}[section]
\newenvironment{example}{\begin{Exa}\em }{\end{Exa}}

\def\Comment{Comment}
\newtheorem{Com}{\Comment}[section]
\newenvironment{remark}{\begin{Com}\em }{\end{Com}}

\def\indbb#1#2
 {\if#11\arabic{#2}\else\if#1a\alph{#2}\else\if#1A\Alph{#2}\else%
  \if#1i\roman{#2}\else\if#1I\Roman{#2}\else%
        {{#1}$_{\arabic{#2}}$}\fi\fi\fi\fi\fi}

\def\widbb#1%
 {\if#11{9}\else\if#1a{a}\else\if#1A{A}\else%
  \if#1i{vi}\else\if#1I{VI}\else{9#1}\fi\fi\fi\fi\fi}

{\begin{list}{\hspace*{3ex}{\rm\bf (\indbb{#1}{enumi})}%
                                          \hspace*{2ex}}%
 {\labelwidth 0pt\leftmargin 0pt
  \topsep 2pt\itemsep 0pt\usecounter{enumi}}}
{\end{list}}

\newenvironment{bbize}[1]
{\if#1*%
\begin{list}{{$\bullet$}}%
 {\settowidth\labelwidth{$\bullet$}%
  \leftmargin\labelwidth\advance\leftmargin\labelsep%
  \topsep 2pt\itemsep 0pt\usecounter{enumi}}%
\else
\begin{list}{{\rm\bf (\indbb{#1}{enumi})}\hfill}%
 {\settowidth\labelwidth{(\widbb{#1})}%
  \leftmargin\labelwidth\advance\leftmargin\labelsep%
  \topsep 2pt\itemsep 0pt\usecounter{enumi}}%
\fi}
{\end{list}}

\newenvironment{abbize}[2]
{\begin{list}{\hspace*{3ex}{\rm\bf (\indbb{#1}{enumi}%
                                    \indbb{#2}{enumii})}%
 }%
 {\labelwidth 0pt\leftmargin 0pt
  \topsep 0pt\itemsep 0pt\usecounter{enumii}}}%
{\end{list}}

\newenvironment{bbbize}[2]%
{\if#2*%
\begin{list}{{\bf --}}%
 {\settowidth\labelwidth{--}%
  \leftmargin\labelwidth\advance\leftmargin\labelsep%
  \topsep 0pt\itemsep 0pt\usecounter{enumii}}%
\else%
\begin{list}{{\rm\bf (\indbb{#1}{enumi}\indbb{#2}{enumii})}\hfill}%
 {\settowidth\labelwidth{(\widbb{#1}\widbb{#2})}%
  \leftmargin\labelwidth\advance\leftmargin\labelsep%
  \topsep 0pt\itemsep 0pt\usecounter{enumii}}%
\fi}%
{\end{list}}

\newcommand{\CLOCK}[9]{
\begin{picture}(0,0)\thicklines%
 \put(0,0){\makebox(0,0)[c]{$\bullet$}}%
 \DBL=#1\multiply\DBL by 7\divide\DBL by 10
 \HALF=#1\multiply\HALF by 2\divide\HALF by 5
 $\bezier{#1}(0,#1)       (\HALF,#1)  (\DBL,\DBL)$%
 $\bezier{#1}(\DBL,\DBL)  (#1,\HALF)  (#1,0)$%
 $\bezier{#1}(#1,0)       (#1,-\HALF) (\DBL,-\DBL)$%
 $\bezier{#1}(\DBL,-\DBL) (\HALF,-#1) ( 0,-#1)$%
 $\bezier{#1}(0,-#1)      (-\HALF,-#1)(-\DBL,-\DBL)$%
 $\bezier{#1}(-\DBL,-\DBL)(-#1,-\HALF)(-#1,0)$%
 $\bezier{#1}(-#1,0)      (-#1, \HALF)(-\DBL,\DBL)$%
 $\bezier{#1}(-\DBL,\DBL) (-\HALF, #1)(0,#1)$%
 \put(0,#1){\line(0,-1){4}}
 \put(#1,0){\line(-1,0){4}}%
 \put(0,-#1){\line(0,1){4}}%
 \put(-#1,0){\line(1,0){4}}%
 \HALF=#1 \advance\HALF by 5%
 \DBL=\HALF\multiply\DBL by 7\divide\DBL by 10
 \put(0,\HALF){\makebox(0,0)[b]{#2}}%
 \put(\DBL,\DBL){\makebox(0,0)[bl]{#3}}%
 \put(\HALF,0){\makebox(0,0)[l]{#4}}%
 \put(\DBL,-\DBL){\makebox(0,0)[tl]{#5}}%
 \put(0,-\HALF){\makebox(0,0)[t]{#6}}%
 \put(-\DBL,-\DBL){\makebox(0,0)[tr]{#7}}%
 \put(-\HALF,0){\makebox(0,0)[r]{#8}}%
 \put(-\DBL,\DBL){\makebox(0,0)[br]{#9}}%
\end{picture}}

\long\def\vokrug#1#2#3#4
{\hbox{\vbox{\hangindent=#1\hangafter=#2{#3}}%
 \kern -\textwidth
 \fbox{%
 \LtoR=#1\advance\LtoR by -1em\parbox[b]{\LtoR}{#4}%
       }
}}

\long\def\rvokrug#1#2#3#4
{\LtoR=#1\advance\LtoR by 1em%
 \hbox{\vbox{\hangindent=\LtoR\hangafter=#2{#4}}%
  \kern -\textwidth
  \LtoR=#1\advance\LtoR by -.7em
  \fbox{\parbox[b]{\LtoR}{\centerline{#3}}}}
 \hangindent=\LtoR\hangafter=-1
}

\long\def\lvokrug#1#2#3#4
{\LtoR=#1\advance\LtoR by 1em%
 \hbox{\vbox{\hangindent=-\LtoR\hangafter=#2{#4}}%
  \if#20\kern 1em\else\kern -#1\fi%
  \LtoR=#1\advance\LtoR by -.7em%
  \fbox{\parbox[b]{\LtoR}{\centerline{#3}}}}
 \hangindent=-\LtoR\hangafter=-1
}

\long\def\answer#1#2{%
 \ifanswers\noindent\underline{{\bf Answer:}} \quad {#1}~%
   {\if#2\else\\ \underline{{\bf Explanation:}}\quad{#2}\fi}%
 \else\noindent{\em Answer\/}
   (\underline{Give a complete explanation}):\vspace*{\fill}\fi}

\newcounter{pointscounter}

   \ifmy\input{ddate.tex}\fi

\def\bang#1{\mbox{{\rm\ !}\(#1\)}}

\newcommand{ \epH }{\mbox{{\rm Horn }}}
\long\def\comment#1{}
\long\def\done#1#2{\ifmy{%
 \if#11{\ifpdone{\footnote{{\large\bf begin-done}}%
             {#2}\footnote{{\large\bf end-done}}}\fi}
            \else\ifpdone\footnote{{\huge\bf do-it}\ {#2}}\fi\fi}\fi}

\long\def\done#1#2{\ifmy{%
 \if#10{\ifpdone~\\===\footnote{{\huge\bf do-it}\ {#2}}\fi}\fi}\fi}

\def\good#1{{\mbox{\(\displaystyle{#1}\)}}}
\def\pH{\mbox{$\oplus$-Horn}\ }

\def\epH{\mbox{{\rm Horn}\ }}
\def\OUT#1{\good{\mbox{{\rm VALUE}}(#1)}}

\def\SouthEast#1#2#3
{\begin{picture}(0,0)\thicklines
 \put(0,0){\vector(1,-1){#1}}
 \HALF=#1 \divide\HALF by 2
 \put(\HALF,-\HALF){\makebox(0,0)[b#3]{{#2}}}
\end{picture}}

\def\SouthWest#1#2#3
{\begin{picture}(0,0)\thicklines
 \put(0,0){\vector(-1,-1){#1}}
 \HALF=#1 \divide\HALF by 2
 \put(-\HALF,-\HALF){\makebox(0,0)[b#3]{{#2}}}
\end{picture}}

\def\South#1#2#3
{\begin{picture}(0,0)\thicklines%
 \put(0,0){\vector(0,-1){#1}}
 \HALF=#1 \divide\HALF by 2
 \put(0,-\HALF){\makebox(0,0)[b#3]{\underline{#2}}}
\end{picture}}

\def\SouthD#1
{\begin{picture}(0,0)\thicklines%
 \TreeH=#1 \divide\TreeH by 4
 \put(0,0){\line(0,-1){\TreeH}}
 \TreeW=#1 \divide\TreeW by 12
 \advance \TreeH by \TreeW
 \multiput(0,-\TreeH)(0,-\TreeW){5}{\makebox(0,0)[c]{$\cdot$}}
 \TreeH=#1 \divide\TreeH by 4 \TreeW=\TreeH \multiply\TreeW by 3
 \put(0,-\TreeW){\vector(0,-1){\TreeH}}
\end{picture}}
\def\SouthEastD#1
{\begin{picture}(0,0)\thicklines%
 \TreeH=#1 \divide\TreeH by 4
 \put(0,0){\line(1,-1){\TreeH}}
 \TreeW=#1 \divide\TreeW by 12
 \advance \TreeH by \TreeW
 \multiput(\TreeH,-\TreeH)(\TreeW,-\TreeW){5}{\makebox(0,0)[c]{$\cdot$}}
 \TreeH=#1 \divide\TreeH by 4 \TreeW=\TreeH \multiply\TreeW by 3
 \put(\TreeW,-\TreeW){\vector(1,-1){\TreeH}}
\end{picture}}

\def\SouthEEast#1#2#3
{\begin{picture}(0,0)\thicklines
\QUA=#1 \multiply \QUA by 4
 \put(0,0){\vector(4,-1){\QUA}}
 \HALF=#1 \divide\HALF by 2\QUA=#1 \multiply \QUA by 2
 \put(\QUA,-\HALF){\makebox(0,0)[b#3]{{#2}}}
\end{picture}}

\def\SouthWWest#1#2#3
{\begin{picture}(0,0)\thicklines
\QUA=#1 \multiply \QUA by 4
 \put(0,0){\vector(-4,-1){\QUA}}
 \HALF=#1 \divide\HALF by 2\QUA=#1 \multiply \QUA by 2
 \put(-\QUA,-\HALF){\makebox(0,0)[b#3]{{#2}}}
\end{picture}}

\def\red#1{{\color{red}{{\boldmath\bf\mbox{#1}}}}}%
\def\rede#1{{\color{red}{{\boldmath\bf{#1}}}}}%
\def\blue#1{{\color{blue}{{\boldmath\bf\mbox{#1}}}}}%
\def\bluee#1{{\color{blue}{{\boldmath\bf{#1}}}}}%
\def\HLL{\mbox{{\rm\bf HLL}}}

\newcommand{ \lconf}[4]{\mbox{$({#1} \otimes (r_1^\good{#2} \otimes
         r_2^\good{#3} \otimes \cdots \otimes r_n^\good{#4}))$}}
\newcommand{ \mconf}[4]{\mbox{$({#1},\ \good{#2},
                    \good{#3}, \ldots, \good{#4})$}}
\newcommand{ \lconfe}[5]{\mbox{$({#1} \otimes (r_1^\good{#2} \otimes
  r_2^\good{#3}\otimes\cdots\otimes r_m^\good{#4}\otimes\cdots\otimes
  r_n^\good{#5}))$}}
\newcommand{ \mconfe}[5]{\mbox{$({#1},\ \good{#2},
  \good{#3}, \ldots, \good{#4}, \ldots, \good{#5})$}}

\begin{document}
\title{{\bf Horn Linear Logic and Minsky Machines}
 \\ ({\tt adapted, reshuffled, self-contained, fully detailed,
 cleansed, simplified,
 with pictures by myself, etc.})}
\author{Max Kanovich\\ University College London, UK and
 \\ National Research University Higher School of Economics}
\date{Twenty Years After} 
\maketitle
\tableofcontents
\listoftables
\listoffigures

\newpage
\begin{abstract}
 Here we give a detailed proof for 
 the crucial point in our Minsky machine simulation:

{\bf Theorem~\ref{t-lLLM}}
{\em Any linear logic derivation for a Horn sequent of the form
$$\seq{\lconf{l_1}{k_1}{k_2}{k_n},\,\bang{\Phi_M},\,
 \bang{{\cal K}}}{l_0}$$%
 can be transformed into a Minsky computation 
 leading from an initial configuration of the form
$$ \mconf{L_1}{k_1}{k_2}{k_n} $$%
 to
 the halting configuration $$\mconf{L_0}{0}{0}{0}.$$
}

 For the sake of perspicuity I include the information
 about the main encoding.
 In particular, this specifies what kind
 of Horn programs of a simple branching structure
 we are actually dealing with within the framework
 of our particular encoding.

Among other things, the presentation advantage
of the 3-step program is that the
non-trivial tricky points are distributed between the
independent parts each of which we justify following
 its own intrinsic methodology
 (to say nothing of the induction used in the opposite directions):

\begin{bbize}{1}
\item
 From LL to HLL -
     we use purely proof-theoretic arguments.
\item From HLL to Horn programs - we translate trees
     (HLL derivations) into another trees (Horn programs)
     of the same shape, almost.
\item From Horn programs to Minsky computations -
     we use purely computational arguments.
\end{bbize}

Since the unavoidable implication of the 3-step program
is undecidability of full linear logic,
 I would highly appreciate your comments on
 which issues looked suspicious to you to be addressed to
 with further detalization.
\end{abstract}

\section{From linear logic to Horn Linear Logic {\bf HLL} }
\label{s-LL-HLL}

 We start from the purely proof-theoretic part:
\begin{theorem}\label{t-ll-HLL}
 Any cut-free derivation for a Horn sequent of the form
      $$\seq{W,\,\Gamma,\,\bang{\Delta}}{Z} $$%
 can be transformed into a derivation in Horn Linear Logic, \HLL. 
\end{theorem}

\subsection{Language: Horn sequents}   
 The connectives $\otimes$ and\/~$\oplus$ are assumed
 to be commutative and associative.

 Here we confine ourselves to {\em Horn-like sequents\/} introduced in
 the following way.

\begin{definition}  \label{dsimple}
 The tensor product of {\em a positive number} of positive literals is
 called {\bf a simple product.}
 A single literal~$q$ is also called {\bf a simple product.}
\end{definition}

\begin{definition}  \label{dmsetprod}
 We will use a natural isomorphism between non-empty finite multisets over
 positive literals and simple products:\ \
 \mbox{A~multiset\ \ $\{q_1,\ q_2, \ \ldots, \ q_k\}$}\ \ \
 is represented by the simple product\ \ \
 \mbox{$(q_1 \otimes q_2 \otimes \cdots \otimes q_k),$}\ \ \
 and vice versa.
\end{definition}

\begin{definition}
 We will write\ \ \ \ \mbox{$ X \cong Y $}\ \ \ \
 to indicate that $X$~and~$Y$ represent one and the same multiset~$M$.
\end{definition}

\begin{definition}
 Here, and henceforth,
 $X$, $X'$, $Y$, $Y_i$, $U$, $V$, $W$, $Z$, etc.,
 stand for {\em \mbox{$\otimes$}-products of positive literals}.

\noindent {\bf Horn implications}\  are defined as follows:
\begin{bbize}{i}
\item {\bf A Horn implication}\  is a formula of the form
         $$\limply{X}{Y}.$$
\item {\bf A~$\oplus$-Horn implication}\  is a formula of the form
         $$\lvariant{X}{Y_1}{Y_2}.$$
\end{bbize}
\end{definition}

\begin{definition}
 A {\bf Horn sequent\/} is defined as a sequent of the form
          $$\seq{W,\Gamma,\bang{\Delta}}{Z}$$%
 where the multisets~$\Gamma$ and~$\Delta$
 consist of Horn implications and $\oplus$-Horn implications,
 and $W$~and\/~$Z$ are {\em \mbox{$\otimes$}-products
 of positive literals.}
\end{definition}

\subsection{%
 LL rules used in cut-free derivations for Horn sequents} 

 In Table~\ref{tLL} we collect all the inference rules of Linear Logic
 that can be used in cut-free derivations for Horn sequents.
\begin{bbize}{i}
\item
 ``Left rules'': {\bf L$\otimes$}, {\bf L$\llto$},
 {\bf L$\llto\oplus$}, {\bf L$\oplus$},
   {\bf L!}, {\bf W!}, {\bf C!}.
\item
 ``Right rules'':
 {\bf R$\otimes$}.
\end{bbize}
 The {\em intuitionistic shape\/}
 of the rules selected in Table~\ref{tLL}
 is caused by the fact that a sequent of the form
      $$\seq{W,\,\Gamma,\,\bang{\Delta}}{} $$%
 is not derivable in linear logic - simply replace all propositions
 with the constant\/~${\boldmath\une}$.
\begin{table*}[htp]
\begin{center}
\begin{tabular}{|rlrl|}     \hline &&&\\
{\bf I}          & \son{}{\seq{X}{X}}  &  &
\\[1em]
{\bf L$\otimes$} & \son{\seq{\Sigma,X,Y}{Z{}}}
                       {\seq{\Sigma,(X \otimes Y)}{Z{}}} &
{\bf R$\otimes$} &
 \son{\seq{\Sigma_1}{Z{}_1} \ \ \ \ \ \seq{\Sigma_2}{Z{}_2}}
     {\seq{\Sigma_1,\Sigma_2}{(Z{}_1\otimes Z{}_2)}}
\\[1em]
{\bf L$\llto$}   &
 \son{\seq{\Sigma_1}{X} \ \ \ \ \ \seq{Y,\ \Sigma_2}{Z{}}}
     {\seq{\Sigma_1,\ \limply{X}{Y},\ \Sigma_2}{Z{}}}    & 
\comment{
{\bf R$\llto$}   &
 \son{\seq{\Sigma,A}{B}}
     {\seq{\Sigma}{\limply{A}{B}}}                     
} 
&
\\[1em]
{\bf L$\llto\oplus$}   &
 \son{\seq{\Sigma_1}{X} \ \ \ \ \
 \seq{(Y_1\oplus Y_2),\ \Sigma_2}{Z{}}}
     {\seq{\Sigma_1,\ \limply{X}{(Y_1\oplus Y_2)},\ \Sigma_2}{Z{}}}
& &
\\[1em]
{\bf L$\oplus$}  &
 \son{\seq{\Sigma,Y_1}{Z{}} \ \ \ \ \seq{\Sigma,Y_2}{Z{}}}
     {\seq{\Sigma,(Y_1 \oplus Y_2)}{Z{}}}                    &
\comment{
{\bf R$\oplus$}   & 
 \son{\seq{\Sigma}{A,Z{}}}
     {\seq{\Sigma}{(A \oplus B),Z{}}}\ 
 \son{\seq{\Sigma}{B,Z{}}}
     {\seq{\Sigma}{(A \oplus B),Z{}}}
} 
& \\[1em]
{\bf L!}         &
 \son{\seq{\Sigma, A}{Z{}}}
     {\seq{\Sigma,\bang{A}}{Z{}}}                        &
\comment{
{\bf R!}         &
 \son{\seq{\bang{\Sigma}}{Z{}}}
     {\seq{\bang{\Sigma}}{\bang{Z{}}}}
} 
 &\\[1em]
{\bf W!}         &
 \son{\seq{\Sigma}{Z{}}}
     {\seq{\Sigma,\bang{A}}{Z{}}}                        &
{\bf C!}         &
 \son{\seq{\Sigma,\bang{A},\bang{A}}{Z{}}}
     {\seq{\Sigma,\bang{A}}{Z{}}}
\\[1em] \hline
\end{tabular}
\end{center}
\caption {The linear logic rules we use for Horn sequents.
  Here $A$ is a Horn implication or \pH implication}
\label{tLL}
\end{table*}

\subsection {The Inference Rules of Horn Linear Logic}

 The inference rules of the Horn Linear Logic
 {\bf HLL} are given in Table~\ref{tHLL}.

\begin{table*}[htp]
\begin{center}
\begin{tabular}{|rlrl|}               \hline &&&\\
{\bf I}          & \son{}{\seq{X}{X}}  &
{\bf L$\otimes$} &
 \son{\seq {X,\Gamma,\bang{\Delta}}{Z}}
     {\seq {Y,\Gamma,\bang{\Delta}}{Z}}(where $X \cong Y$)\\[3ex]
{\bf H}          & \son{}
                       {\seq {X,\limply{X}{Y}}{Y}} &
{\bf M}          &  
\son {\seq{X,\Gamma,\bang{\Delta}}{Y}}
     {\seq{(X \otimes V),\Gamma,\bang{\Delta}}{(Y \otimes V)}}\\[3ex]
{\bf $\oplus$-H}                   &\multicolumn{3}{l|}{
\son {\seq{(Y_1 \otimes V),\Gamma,\bang{\Delta}}{Z} \ \ \ \ \
      \seq{(Y_2 \otimes V),\Gamma,\bang{\Delta}}{Z}}
{\seq{(X \otimes V),\Gamma,\ \lvariant{X}{Y_1}{Y_2},\
                           \bang{\Delta}}{Z}}          }\\[3ex]
{\bf L!}         &
 \son{\seq{X,\Gamma,A,\bang{\Delta}}{Z}}
     {\seq{X,\Gamma,\bang{A},\bang{\Delta}}{Z}}        &&\\[3ex]
{\bf W!}         &
 \son{\seq{X,\Gamma,\bang{\Delta}}{Z}}
     {\seq{X,\Gamma,\bang{A},\bang{\Delta}}{Z}}        &
{\bf C!}         &
 \son{\seq{X,\Gamma,\bang{A},\bang{A},\bang{\Delta}}{Z}}
     {\seq{X,\Gamma,\bang{A},\bang{\Delta}}{Z}}    \\[3ex]
{\bf Cut}                          &\multicolumn{3}{l|}{
\son {\seq{W,\Gamma_1,\bang{\Delta_1}}{U}  \ \ \ \ \
      \seq{U,\Gamma_2,\bang{\Delta_2}}{Z}}
{\seq{W,\Gamma_1,\Gamma_2,\bang{\Delta_1},\bang{\Delta_2}}{Z}}
                                                       }\\[3ex]
\hline
\end{tabular}
\end{center}
\caption {Horn Linear Logic \HLL.\ 
 Both $\otimes$ and\/~$\oplus$ are assumed to be
 commutative and associative.}
\label{tHLL}
\end{table*}

\subsection{The proof of Theorem~\ref{t-ll-HLL}:
 \ From LL derivations to \HLL\ derivations.}

 Given a cut-free derivation for the sequent
      $$\seq{W,\,\Gamma,\,\bang{\Delta}}{Z} $$%
 by induction we will simulate each of        
 the LL rules in Table~\ref{tLL} with
 the \HLL\ rules from Table~\ref{tHLL}.


\subsection*{Rule {\bf L$\llto$} and the like}  

\begin{bbize}{i}
\item
 An ({\bf R$\otimes$})-rule of the form
 (here $\pi_1$ and~$\pi_2$ are proofs that have been already
  constructed by induction with rules from Table~\ref{tHLL}): 
$$\sonN{\,\blue{{\bf R$\otimes$}}}
    {\sonNl{}{$\pi_1$}{$\Sigma_1\vdash Z_1$}%
     \hspace*{10em}%
     \sonNr{}{$\pi_2$}{$\Sigma_2\vdash Z_2$}%
   }{$\Sigma_1,\Sigma_2\vdash (Z_1\otimes Z_2)$}$$%
 is simulated with the \HLL\ rules from Table~\ref{tHLL}:
$$\hspace*{-14ex}
 \sonN{\,\red{{\bf Cut}}}
   {\sonNl{\,\red{{\bf M}}}
     {\sonNc{}{$\pi_1$}{$\Sigma_1\vdash Z_1$}%
     }{$Z_2, \Sigma_1\vdash (Z_1\otimes Z_2)$}
     \hspace*{30ex}
     \sonNr{}{$\pi_2$}{$\Sigma_2\vdash Z_2$}%
   }{$\Sigma_1,\Sigma_2\vdash (Z_1\otimes Z_2)$}$$%
\item
 An ({\bf L$\llto$})-rule of the form
 (here $\pi_1$ and~$\pi_2$ are proofs that have been already
  constructed by induction with rules from Table~\ref{tHLL}): 
$$\sonN{\,\blue{{\bf L$\llto$}}}
    {\sonNl{}{$\pi_1$}{$\Sigma'\vdash X$}%
     \hspace*{10em}%
     \sonNr{}{$\pi_2$}{$Y,\Sigma''\vdash Z'$}%
    }{$\Sigma', (X\llto Y), \Sigma''\vdash Z'$} $$%
 is simulated with the \HLL\ rules from Table~\ref{tHLL}:
$$\hspace*{-14ex}
 \sonN{\,\red{{\bf Cut}}}
   {\sonNl{\,\red{{\bf Cut}}}
     {\sonNl{}{$\pi_1$}{$\Sigma'\vdash X$}%
      \hspace*{20ex}
      \sonNr{\,\red{{\bf H}}}{}{$X,\, (X\llto Y) \vdash Y$}%
     }{$\Sigma', (X\llto Y)\vdash Y$}
     \hspace*{30ex}
     \sonNr{}{$\pi_2$}{$Y, \Sigma''\vdash Z'$}%
   }{$\Sigma', \limply{X}{Y},\Sigma''\vdash Z'$}$$%
\item
 The remaining LL rules,
 save for {\bf L$\oplus$} and\/ {\bf L$\llto\oplus$},
 are processed by the same token.
\QED
\end{bbize}
\comment{
$$ \sonN{\,{\bf Cut}}
     {\sonNl{\,{\bf M}}
        {\sonNcc{}{$\pi_1$}{$X',\Sigma_1,\bang{\Delta_1}\vdash X$}
        }{$(X'\otimes V),\Sigma_1,\bang{\Delta_1}\vdash(X\otimes V)$}%
      \hspace*{16em}%
      \sonNr{\,{\bf H}}
        {\sonNcc{}{$\pi_2$}
                {$(Y\otimes V),\Sigma_2,\bang{\Delta_2}\vdash Z'$}%
        }{$(X\otimes V),
             \Sigma_2,(X\llto Y),\bang{\Delta_2}\vdash Z'$}%
     }{$(X'\otimes V),\Sigma_1,\Sigma_2,(X\llto Y),
      \bang{\Delta_1},\bang{\Delta_2}\vdash Z'$} $$
} 

\subsection*{Challenging {\bf L$\oplus$} and\/ {\bf L$\llto\oplus$}  }

 The main difficulties we meet with the rule
 {\bf L$\oplus$} (and related to it\/ {\bf L$\llto\oplus$}) are
 that the positions at which these rules are applied in the given
 cut-free LL derivation might have happened very far from
 each other.

\noindent
 First, we have to contract the distance between their positions
 by pushing  {\bf L$\oplus$}  downwards in accordance with
 Lemma~\ref{l-L-oplus} to make the application positions
 of {\bf L$\oplus$} and\/ {\bf L$\llto\oplus$} {\em adjacent\/}:
%
\begin{lemma}\label{l-L-oplus}
 Given a cut-free derivation with the rules from Table~\ref{tLL}, 
 by the appropriate {\bf `commuting conversions'\/} (see below),
 the left rule {\bf L$\oplus$} can be pushed downwards
 (down to the related\/~{\bf L}$\llto\oplus$), forming a piece
 of the derivation where
 the rules {\bf L$\oplus$} and\/ {\bf L$\llto\oplus$} 
 are sitting in the adjacent positions so that
 the rule\/ {\bf L$\llto\oplus$} is applied\/ {\bf just after\/}
 the rule\/ {\bf L$\oplus$}:
\begin{equation}%
     \label{eq-L-oplus}
\hspace*{-18ex}
\sonN{\,\red{{\bf L}$\llto\oplus$}}
   {\sonNl{}{$\pi_0$}{$\Sigma'\vdash X$}
     \hspace*{12em}
    \sonNr{\,\red{{\bf L}$\oplus$}}
     {\sonNl{}{$\pi_1$}{$Y_1,\Sigma''\vdash Z'$}%
      \hspace*{8em}
      \sonNr{}{$\pi_2$}{$Y_2,\Sigma''\vdash Z'$}%
     }{$(Y_1\oplus Y_2),\ \Sigma''\vdash Z'$}%
   }{$\Sigma',\ (X\llto(Y_1\oplus Y_2)),\ \Sigma''\vdash Z'$}%
\end{equation}
\end{lemma}
\noindent{\bf Proof.}
 We consider all points of interaction between
 the rule {\bf L}$\oplus$ and other rules.
\begin{bbize}{a}
\item
 A combination:
 ``first {\bf L}$\oplus$, then {\bf L}$\llto$,'' 
 of the form (here $\pi_0$, $\pi_1$ and\/~$\pi_2$ are proofs):
$$\hspace*{-14ex}
 \sonN{\,\blue{{\bf L}$\llto$}}
   {\sonNl{\,\blue{{\bf L}$\oplus$}}
     {\sonNl{}{$\pi_1$}{$\Sigma',Y_1\vdash U$}%
      \hspace*{8em}
      \sonNr{}{$\pi_2$}{$\Sigma',Y_2\vdash U$}%
     }{$\Sigma', (Y_1\!\oplus\!Y_2)\vdash U$}
     \hspace*{12em}
     \sonNr{}{$\pi_0$}{$V,\Sigma''\vdash Z'$}%
   }{$\Sigma', (Y_1\!\oplus\!Y_2), \limply{U}{V},\Sigma''\vdash Z'$}$$%
 can be replaced with the following combination:
 ``first {\bf L}$\llto$, then {\bf L}$\oplus$:'' 
$$\hspace*{-18ex}\sonN{\,\red{{\bf L}$\oplus$}}
   {\sonNl{\,\red{{\bf L}$\llto$}}
     {\sonNl{}{$\pi_1$}{$\Sigma',Y_1\vdash U$}%
      \hspace*{6em}
      \sonNr{}{$\pi_0$}{$V,\Sigma''\vdash Z'$}%
     }{$\Sigma', Y_1, \limply{U}{V}, \Sigma''\vdash Z'$}%
     \hspace*{16em}
    \sonNr{\,\red{{\bf L}$\llto$}}
     {\sonNl{}{$\pi_2$}{$\Sigma',Y_2\vdash U$}%
      \hspace*{6em}
      \sonNr{}{$\pi_0$}{$V,\Sigma''\vdash Z'$}%
     }{$\Sigma', Y_2, \limply{U}{V}, \Sigma''\vdash Z'$}%
   }{$\Sigma', (Y_1\!\oplus\!Y_2), \limply{U}{V},\Sigma''\vdash Z'$}$$%

\comment{
\item
 A combination:
 ``first {\bf L}$\oplus$, then {\bf L}$\llto$,'' 
 of the form (here $\pi_0$, $\pi_1$ and\/~$\pi_2$ are proofs):
$$\hspace*{-14ex}
 \sonN{\,\blue{{\bf L}$\llto$}}
   {\sonNl{\,\blue{{\bf L}$\oplus$}}
     {\sonNl{}{$\pi_1$}{$V,\Sigma'',Y_1\vdash Z'$}%
      \hspace*{8em}
      \sonNr{}{$\pi_2$}{$V,\Sigma'',Y_2\vdash Z'$}%
     }{$V,\Sigma'', (Y_1\!\oplus\!Y_2)\vdash Z'$}
     \hspace*{14em}
     \sonNr{}{$\pi_0$}{$\Sigma'\vdash U$}%
   }{$\Sigma',\limply{U}{V},\Sigma'',(Y_1\!\oplus\!Y_2),\vdash Z'$}$$%
 can be replaced with the following combination:
 ``first {\bf L}$\llto$, then {\bf L}$\oplus$:'' 
$$\hspace*{-18ex}\sonN{\,\red{{\bf L}$\oplus$}}
   {\sonNl{\,\red{{\bf L}$\llto$}}
     {\sonNl{}{$\pi_1$}{$V,\Sigma'',Y_1\vdash Z'$}%
      \hspace*{6em}
      \sonNr{}{$\pi_0$}{$\Sigma'\vdash U$}%
     }{$\Sigma', \limply{U}{V}, \Sigma'', Y_1\vdash Z'$}%
     \hspace*{14em}
    \sonNr{\,\red{{\bf L}$\llto$}}
     {\sonNl{}{$\pi_2$}{$V,\Sigma'',Y_2\vdash Z'$}%
      \hspace*{6em}
      \sonNr{}{$\pi_0$}{$\Sigma'\vdash U$}%
     }{$\Sigma', \limply{U}{V}, \Sigma'', Y_2\vdash Z'$}%
  }{$\Sigma',\limply{U}{V},\Sigma'', (Y_1\!\oplus\!Y_2),\vdash Z'$}$$%

} 
\item
 A combination:
 ``first {\bf L}$\oplus$, then {\bf R}$\otimes$,'' 
 of the form (here $\pi_0$, $\pi_1$ and~$\pi_2$ are proofs):
$$\hspace*{-18ex}
 \sonN{\,\blue{{\bf R}$\otimes$}}
   {\sonNl{\,\blue{{\bf L}$\oplus$}}
     {\sonNl{}{$\pi_1$}{$\Sigma',Y_1\vdash Z'$}%
      \hspace*{6em}
      \sonNr{}{$\pi_2$}{$\Sigma',Y_2\vdash Z'$}%
     }{$\Sigma', (Y_1\!\oplus\!Y_2)\vdash Z'$}
     \hspace*{10em}
     \sonNr{}{$\pi_0$}{$\Sigma''\vdash Z''$}%
   }{$\Sigma', (Y_1\!\oplus\!Y_2), \Sigma''\vdash (Z'\otimes Z'')$}$$%
 can be replaced with the following combination:
 ``first {\bf R}$\otimes$, then {\bf L}$\oplus$:'' 
$$\hspace*{-20ex}\sonN{\,\red{{\bf L}$\oplus$}}
   {\sonNl{\,\red{{\bf R}$\otimes$}}
     {\sonNl{}{$\pi_1$}{$\Sigma',Y_1\vdash Z'$}%
      \hspace*{6em}
      \sonNr{}{$\pi_0$}{$\Sigma''\vdash Z''$}%
     }{$\Sigma', Y_1,\Sigma''\vdash (Z'\otimes Z'')$}%
     \hspace*{14em}
    \sonNr{\,\red{{\bf R}$\otimes$}}
     {\sonNl{}{$\pi_2$}{$\Sigma',Y_2\vdash Z'$}%
      \hspace*{6em}
      \sonNr{}{$\pi_0$}{$\Sigma''\vdash Z''$}%
     }{$\Sigma', Y_2, \Sigma''\vdash (Z'\otimes Z'')$}%
   }{$\Sigma', (Y_1\!\oplus\!Y_2), \Sigma''\vdash (Z'\otimes Z'')$} $$

\comment{
\item
 A combination:
 ``first {\bf L}$\llto$, then {\bf R}$\otimes$,'' 
 of the form (here $\pi_0$, $\pi_1$ and~$\pi_2$ are proofs):
$$\sonN{\,{\bf R}$\otimes$}
   {\sonNl{\,{\bf L}$\llto$}
     {\sonNl{}{$\pi_1$}{$\Sigma_1\vdash X$}%
      \hspace*{6em}
      \sonNr{}{$\pi_2$}{$Y, \Sigma_2\vdash Z'$}%
     }{$\Sigma_1, (X\llto Y),\Sigma_2\vdash Z'$}
     \hspace*{10em}
     \sonNr{}{$\pi_0$}{$\Sigma''\vdash Z''$}%
   }{$\Sigma_1,(X\llto Y),\Sigma_2,\Sigma''\vdash(Z'\otimes Z'')$} $$
 can be replaced with the following combination:
 ``first {\bf R}$\otimes$, then {\bf L}$\llto$:'' 
$$\sonN{\,{\bf L}$\llto$}
   {\sonNl{}{$\pi_1$}{$\Sigma_1\vdash X$}
     \hspace*{10em}
    \sonNr{\,{\bf R}$\otimes$}
     {\sonNl{}{$\pi_2$}{$Y,\Sigma_2\vdash Z'$}%
      \hspace*{6em}
      \sonNr{}{$\pi_0$}{$\Sigma''\vdash Z''$}%
     }{$Y,\Sigma_2, \Sigma''\vdash (Z'\otimes Z'')$}%
   }{$\Sigma_1,(X\llto Y),\Sigma_2,\Sigma''\vdash(Z'\otimes Z'')$} $$
} 
\item
 The appropriate {\em `commuting conversions'\/}
 for the remaining combinations:
 ``first {\bf L}$\oplus$, then \dots''
 can be constructed in a similar way.
\QED
\end{bbize}

\subsection*{Completing {\bf L$\oplus$} and\/ {\bf L$\llto\oplus$}}

 According to Lemma~\ref{l-L-oplus},
 in order to complete the proof of Theorem~\ref{t-ll-HLL},
 it suffices to take a piece of the derivation where
 the rules {\bf L$\oplus$} and\/ {\bf L$\llto\oplus$} 
 are sitting in the adjacent positions so that
 the rule\/ {\bf L$\llto\oplus$} is applied\/ {\bf just after\/}
 the rule\/ {\bf L$\oplus$}:
$$\hspace*{-18ex}
\sonN{\,\blue{{\bf L}$\llto\oplus$}}
   {\sonNl{}{$\pi_0$}{$\Sigma'\vdash X$}
     \hspace*{12em}
    \sonNr{\,\blue{{\bf L}$\oplus$}}
     {\sonNl{}{$\pi_1$}{$Y_1,\Sigma''\vdash Z'$}%
      \hspace*{8em}
      \sonNr{}{$\pi_2$}{$Y_2,\Sigma''\vdash Z'$}%
     }{$(Y_1\oplus Y_2),\ \Sigma''\vdash Z'$}%
   }{$\Sigma',\ (X\llto(Y_1\oplus Y_2)),\ \Sigma''\vdash Z'$}%
$$%
 and simulate it with the \HLL\ rules from Table~\ref{tHLL}
 as follows:\footnote{
  Here $\Sigma''$ represents a multiset of the form
  \mbox{$V, \Gamma, !\Delta$}, that is
\quad \mbox{$ \Sigma'' = V, \Gamma, !\Delta $}%
\\
 Recall also our convention: \ \ \mbox{$(U\otimes V) = U,V $}%
 } 
$$\hspace*{-18ex}
\sonN{\,\red{{\bf Cut}}}
   {\sonNl{}{$\pi_0$}{$\Sigma'\vdash X$}
     \hspace*{12em}
    \sonNr{\,\red{{\bf $\oplus$-H}}}
     {\sonNl{}{$\pi_1$}{$Y_1,\Sigma''\vdash Z'$}%
      \hspace*{8em}
      \sonNr{}{$\pi_2$}{$Y_2,\Sigma''\vdash Z'$}%
     }{$X,\ (X\llto (Y_1\oplus Y_2)),\ \Sigma''\vdash Z'$}%
   }{$\Sigma',\ (X\llto (Y_1\oplus Y_2)),\ \Sigma''\vdash Z'$}%
$$%
\QED

\section{From {\bf HLL} to tree-like Horn programs}
\label{s-HLL-programs}

 As computational counterparts of Horn sequents, we will consider
 {\em tree-like Horn programs\/} with the following peculiarities:

\begin{definition}
 A {\bf tree-like Horn program\/} is a rooted binary tree such that
\begin{bbize}{a}
\item Every edge of it is labelled by a Horn implication of the form
      \limply{X}{Y}.
\item The root of the tree is specified as the {\bf input} vertex.
      A~terminal vertex, i.e.~a~vertex with no outgoing edges, will
      be specified as an {\bf output} one.
\item A~vertex~$v$ with exactly two outgoing edges~$(v, w_1)$
      and~$(v, w_2)$ will be called {\bf divergent.}
      These two outgoing edges should be labelled by Horn implications
      with one and the same antecedent, say \limply{X}{Y_1} and
      \limply{X}{Y_2}, respectively.
\end{bbize}
\end{definition}

 Now, we should explain how such a program~$P$ runs
 for a given input~$W$.

\begin{definition} \label{dstrong}
  For a given tree-like Horn program~$P$ and any simple product~$W$,
  a {\bf strong computation\/} is defined by induction as follows:

\noindent
  We~assign a simple product~\footnote%
  { This \OUT{P,W,v} is perceived as the intermediate value of the
    {\em strong computation\/} performed by~$P$, which is obtained at
    point~$v$.}
              $$ \OUT{P,W,v} $$
 to each vertex~$v$ of~$P$ in such a way that
\begin{bbize}{a}
\item For the root~$v_0$,\ \ \ \bluee{$$ \OUT{P,W,v_0} = W.$$}
\item For every non-terminal vertex~$v$ and its son~$w$ with the
      edge~$(v, w)$ labelled by a Horn implication~\limply{X}{Y},
      if \OUT{P,W,v} is defined and, for some simple product~$V$:
          $$ \OUT{P,W,v} \cong (X \otimes V),$$%
      then
          $$ \OUT{P,W,w} = (Y \otimes V).$$%
      Otherwise, \OUT{P,W,w} is declared to be undefined.
\end{bbize}
\end{definition}

\begin{definition}
  For a tree-like Horn program~$P$ and a simple product~$W$,
  we say that
               $$ P(W) = Z $$
  if \rede{for each terminal vertex~$w$ of~$P$,
 \OUT{P,W,w}
  is defined and $$\OUT{P,W,w} \cong Z.$$}%
\end{definition}

 We will describe each of our program constructs
 by Linear Logic formulas.
 Namely, we will associate a certain formula~$A$ to each edge~$e$ of a
 given program~$P$, and say that\\
\centerline{{\em ``This formula~$A$ is used on the edge~$e$.''}}

\begin{definition} Let $P$ be a tree-like Horn program.
\begin{bbize}{a}
\item If $v$ is a non-divergent vertex of~$P$
 with the outgoing edge~$e$
      labelled by a Horn implication~$A$, then we will say that\\
\centerline{{\em ``Formula~$A$ itself is used on the edge~$e$.''}}
\item Let $v$ be a divergent vertex of~$P$ with two outgoing
      edges~$e_1$ and~$e_2$ labelled by Horn implications
      \limply{X}{Y_1} and \limply{X}{Y_2}, respectively.
      Then we will say that\\
        \centerline{{\em ``Formula~$A$ is used on~$e_1$.''}}\\
      and\\
        \centerline{{\em ``Formula~$A$ is used on~$e_2$.''}}\\
      where formula~$A$ is defined as the following \pH implication:
             $$ A = \lvariant{X}{Y_1}{Y_2}.$$
\end{bbize}
\end{definition}

\begin{definition}
 A tree-like Horn program~$P$ is said to be a
 {\bf strong solution to\/} a \epH sequent of the form
     $$ \seq{W,\,\Delta,\,\bang{\Gamma}}{Z} $$
  if \bluee{for each terminal vertex~$w$} of~$P$,
 \OUT{P,W,w}
  is defined and \rede{$$\OUT{P,W,w} \cong Z.$$}%
  and
\begin{bbize}{a}
\item For every edge~$e$ in~$P$, the formula~$A$ used on~$e$ is drawn
       either from~$\Gamma$ or from~$\Delta$.
\item Whatever path~$b$ leading from the root to a terminal vertex
 we take,
 each formula~$A$ from~$\Delta$ is used once and {\bf exactly} once
 on this path~$b$.
\end{bbize}
\end{definition}


\comment{
\begin{definition}
  For a Horn program~$P$ and a simple product~$W$,
  we say that
          $$ P(W) = Z $$
  if and only if for each output vertex~$v$ of~$P$, \OUT{P,W,v}
  is defined and
          $$ \OUT{P,W,v} = Z.$$
\end{definition}
    These definitions fall within the paradigm of Linear Logic,
 ensuring that
\begin{bbize}{a}
 \item the execution of a Horn program does not allow for its
       operators to share their inputs,
 \item after the program has been executed, the memory that was
       occupied by temporary and auxiliary objects is free.
\end{bbize}
} 

\comment{
\begin{example} {\em (continuing Example~\ref{eEX})}
      
 Program~$P_0$ (see Figure~\ref{fEX}) is a strong solution to
 our sequent
   $$\seq {f,\bang{\lvariant{f}{g}{h}}, \bang{\limply{g}{m}},
             \bang{\limply{h}{m}}} {m}.$$
 The corresponding formula indicator~$u$ can be defined by the
 table:
$$\begin{array}{|l|l|} \hline
 \mbox{Edge}\  e & \mbox{Formula}\  u(e) \\ \hline
  (v_0,v_1) & \lvariant{f}{g}{h} \\
  (v_0,v_2) & \lvariant{f}{g}{h} \\
  (v_1,v_3) & \limply{g}{m} \\
  (v_2,v_4) & \limply{h}{m} \\     \hline
\end{array}$$
\end{example}
} 

\def\Node#1#2
{\begin{picture}(0,0)
  \put(0,0){\makebox(0,0)[c]{$\bullet$}}
  \put(0,0){\makebox(0,0)[b#2]{\underline{#1}}}
\end{picture} }

\def\NodeXIV#1#2#3
{\begin{picture}(0,0)
  \put(-7,-14)       {\framebox(14,14){#1}  }
  \if#3r \put(-7,-7) {\makebox(0,0)[br]{\underline{#2}}}%
  \else  \put( 7,-7) {\makebox(0,0)[bl]{\underline{#2}}}\fi
\end{picture} }

\def\Tree#1#2#3
{\begin{picture}(0,0)\thicklines%
 \TreeH=#1 \multiply \TreeH by 2
 \TreeW=#1 \divide \TreeW by 2
 \put(0,0){\line(-1,-4){\TreeW}}
 \put(0,0){\line( 1,-4){\TreeW}}
 \put(0,-#1){\makebox(0,0)[t]{#2}}
 \put(-\TreeW,-\TreeH){\line(1,0){#1}}
 \put(-\TreeW,-\TreeH){\vector( 0,-1){18}}                
 \put( \TreeW,-\TreeH){\vector( 0,-1){18}}                
 \advance \TreeH by 4
 \put(0,-\TreeH){\makebox(0,0)[t]{{\bf \ldots}}}
 \advance \TreeH by 14                             
 \put(-\TreeW,-\TreeH){\Node{#3}{\toRight}}
 \put( \TreeW,-\TreeH){\Node{#3}{\toRight}}
\end{picture}}

\def\TreeWide#1#2#3
{\begin{picture}(0,0)\thicklines%
 \TreeH=#1\multiply\TreeH by 2%
 \TreeW=#1\multiply\TreeW by 5\divide\TreeW by 2
 \put(0,0){\line(-3,-1){\TreeW}}
 \put(0,0){\line( 3,-1){\TreeW}}
 \HALF=#1\divide\HALF by 3%
 \put(0,-\HALF){\makebox(0,0)[t]{#2}}
 \HALF=#1\advance\HALF by -8%
 \put(0,-\HALF){\line(-1,0){\TreeW}}
 \put(0,-\HALF){\line( 1,0){\TreeW}}
 \put(-\TreeW,-\HALF){\vector( 0,-1){18}}                
 \put( \TreeW,-\HALF){\vector( 0,-1){18}}                
 \put(      0,-\HALF){\vector( 0,-1){18}}                
 \advance \HALF by 4
 \divide\TreeW by 2
 \put(-\TreeW,-\HALF){\makebox(0,0)[t]{{\bf \ldots}}}
 \put( \TreeW,-\HALF){\makebox(0,0)[t]{{\bf \ldots}}}
 \advance \HALF by 14                             
 \multiply\TreeW by 2
 \put(-\TreeW,-\HALF){\Node{#3}{\toRight}}
 \put( \TreeW,-\HALF){\Node{#3}{\toRight}}
 \put( 0,-\HALF){\Node{#3}{\toRight}}
\end{picture}}

\hyphenation {de-ter-mi-nistic in-de-ter-mi-nist-ical-ly}

\newcommand{ \Id }[1]%
{\begin{picture}(0,0)  \thicklines
 \Ycur=14 \advance \Ycur by \HALF
 \put(0,-\Ycur){\Node{\ X}{\toRight}}
\end{picture}}

\newcommand{ \IMPLY }[1]%
{\begin{picture}(0,0)  \thicklines
 \Ycur=14
 \put(0,-\Ycur){\NodeXIV{$v$}{$\ \ X$}{\toRight}}
 \advance \Ycur by 14
 \put(0,-\Ycur){\South {#1} {\ \limply{X}{Y}}{\toRight}}
 \advance \Ycur by #1
 \put(0,-\Ycur){\NodeXIV{$w$}{$\ \ Y$}{\toRight}}

\comment{
 \put(0,-\Ycur){\Node{\ X}{\toRight}}
 \put(0,-\Ycur){\South {#1}{\,\limply{X}{Y}}{\toRight}}
 \advance \Ycur by #1
 \put(0,-\Ycur){\Node{\ Y}{\toRight}}
} 
\end{picture}}

 We prove that the  Horn fragment of Linear Logic
 {\bf is complete under our computational interpretation.}

\begin{theorem}[Fairness]   \label{tlinprog}
 Given an \HLL\ derivation (with the rules from Table~\ref{tHLL})
 for a \epH sequent of the form
   $$ \seq{W,\,\Delta,\,\bang{\Gamma}}{Z},$$%
 we can construct a tree-like Horn program\/~$P$
 which is a strong solution to the given sequent.
\end{theorem}

\noindent{\bf Proof.}
 For a given \HLL\ derivation, running from its leaves (axioms)
 to its root, we assemble a tree-like Horn program\/~$P$ by induction.

\noindent
 Below we consider all cases
 related to the rules from Table~\ref{tHLL}.

\begin{figure}[htp]
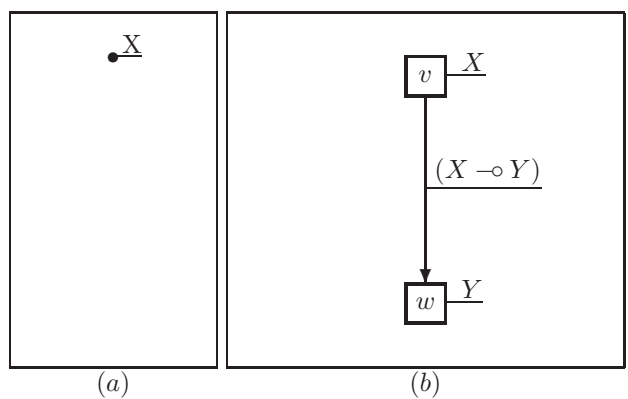

\begin{center}
\( \stackrel {\fbox{\vPICTURE {\Id} {1}{4}{1} {72}}} {(a)} \)
\( \stackrel {\fbox{\vPICTURE {\IMPLY} {1}{4}{2} {72}}} {(b)} \)
\end{center}
\caption { Elementary Horn Programs.}
\label{fIMPLY}
\end{figure}

 {\bf Case~of~Rule~I.} The elementary program from
 Figure~\ref{fIMPLY}(a), with its single vertex, will be a
 strong solution to any sequent of the form
           $$ \seq{X}{X}. $$

 {\bf Case~of~Rule~H.} The elementary program consisting of a
 single edge labelled by \limply{X}{Y} (see~Figure~\ref{fIMPLY}(b))
 will be a strong solution to the sequent
          $$ \seq{X,\,\limply{X}{Y}}{Y}. $$

\newcommand{\M}[1]%
{\begin{picture}(0,0) \thicklines
 \Ycur=14
 \TreeW=#1\multiply\TreeW by 2
 \put(-\TreeW,-\Ycur){\NodeXIV{$v_0$}{\ \ $X$}{\toRight}}
 \put(\TreeW,-\Ycur){\NodeXIV{$v_0$}{\ \ $(X \otimes V)$}{\toRight}}
 \advance \Ycur by 14
\DBL=#1\multiply\DBL by 3\divide\DBL by 2
 \put(-\TreeW,-\Ycur){\Tree{\DBL}{$P$}{$\ Y$}}
 \put(\TreeW,-\Ycur){\Tree{\DBL}{$P'$}{$\,(Y\otimes V)$}}
 \advance \Ycur by \TreeW
 \put(0,-\Ycur){\makebox(0,0)[c]{$\Longrightarrow$}}
\end{picture}}

\begin{figure*}[htp]
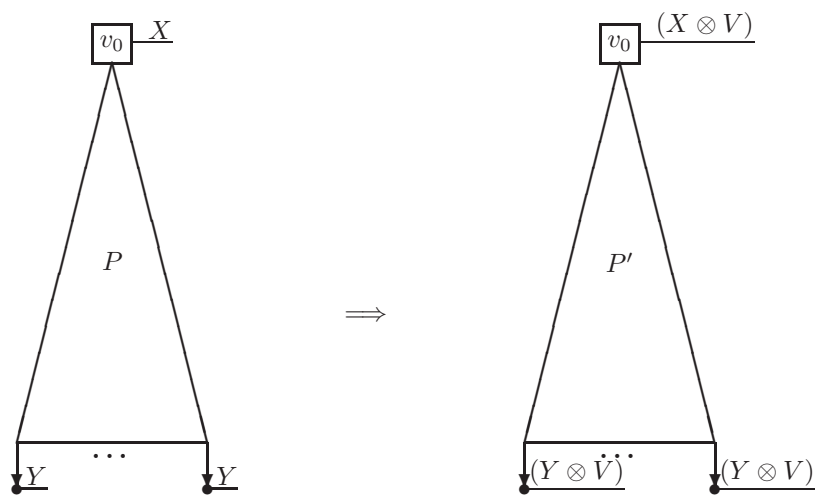

\begin{center}  \vPICTURE {\M} {3}{3}{4} {48}
\end{center}
\caption {The ``Frame property''}
\label{fM}
\end{figure*}


 {\bf Case~of~Rule~M.}    
 Suppose that $P$, with the input\/~$X$,
 is a strong solution to a sequent of the form
 $$\seq{X,\Gamma,\bang{\Delta}}{Y} $$%
 Then as a Horn program\/~$P'$ we take the same~$P$ but with
 a larger input\/~\blue{$(X \otimes V)$}, so that, for any vertex~$w$\
 (see~Figure~\ref{fM}):
\bluee{$$ \OUT{P',(X \otimes V),w} = (\OUT{P,X,w}\otimes V).$$}%
 It is easily verified this Horn program~$P'$ is a strong
 solution to the sequent
 $$  {\seq{(X \otimes V),\Gamma,\bang{\Delta}}{(Y \otimes V)}} $$%

\newcommand{\TwoLOW}[1]%
{\begin{picture}(0,0) \thicklines
 \Ycur=14
 \put(0,-\Ycur){\NodeXIV{$v_0$}{\ \ $(X \otimes V)$}{\toRight}}

 \advance \Ycur by 14
 \put(0,-\Ycur){\SouthWest{#1}{\limply{X}{Y_1}\ }{\toLeft}}
 \put(0,-\Ycur){\SouthEast{#1}{\ \limply{X}{Y_2}}{\toRight}}

 \advance \Ycur by #1
 \put(-#1,-\Ycur){\NodeXIV{$v_1$}{$(Y_1 \otimes V)$\ }{\toLeft}}
 \put( #1,-\Ycur){\NodeXIV{$v_2$}{\ $(Y_2 \otimes V)$}{\toRight}}

 \advance \Ycur by 14
 \put(-#1,-\Ycur){\Tree{#1}{$P_1$}{$\,Z$}}
 \put( #1,-\Ycur){\Tree{#1}{$P_2$}{$\,Z$}}
\end{picture}}

\begin{figure*}[htp]
\begin{center}  \vPICTURE {\TwoLOW} {3}{4}{4} {48}
\end{center}
\caption {Strong Forking. An~\pH implication\
 \mbox{$\limply{X}{(Y_1\oplus Y_2)}$}\ as Non-deterministic choice}
\label{fTwoLOW}
\end{figure*}

 {\bf Case~of~Rule~$\oplus$-H.}
 Suppose that  $P_1$ and~$P_2$ are strong solutions to sequents
 of the form
     $$ \seq{(Y_1 \otimes V),\,\Gamma,\,\bang{\Delta}}{Z} $$
 and $$ \seq{(Y_2 \otimes V),\,\Gamma,\,\bang{\Delta}}{Z}, $$
 respectively.

 Now a Horn program\/~$P'$ can be assembled with the help of
 the following operation of {\bf Strong Forking}\ 
 (see~Figure~\ref{fTwoLOW}):
\begin{bbize}{a}
\item First, we create a new input vertex~$v_0$.
\item After that, we connect\/ $v_0$ with the roots
      $v_1$ of~$P_1$ and\/ $v_2$ of~$P_2$,
      and label the edge~$(v_0,v_1)$ by the
 Horn implication \limply{X}{Y_1}
      and label the edge~$(v_0,v_2)$ by the
 Horn implication \limply{X}{Y_2}.
\end{bbize}

 It is easily verified this Horn program~$P'$ is a strong
 solution to the sequent
 $$ \seq{(X \otimes V),\Gamma,\ \lvariant{X}{Y_1}{Y_2},\
                       \bang{\Delta}}{Z}.$$
\newcommand{ \LOW }[1]%
{\begin{picture}(0,0) \thicklines
 \Ycur=14
 \put(0,-\Ycur){\NodeXIV{$v$}{$\ \ W$}{\toRight}}
 \advance \Ycur by 14
 \put(0,-\Ycur){\TreeWide {#1}{$P_1$}{\ $U$}}
 \advance \Ycur by #1
 \advance \Ycur by 10
 \Xcur=#1\multiply\Xcur by 5\divide\Xcur by 2
 \put(-\Xcur,-\Ycur){\Tree{#1}{$P_2$}{$\,Z$}}
 \put( 0,-\Ycur){\Tree{#1}{$P_2$}{$\,Z$}}
 \put( \Xcur,-\Ycur){\Tree{#1}{$P_2$}{$\,Z$}}
\end{picture}}     

\begin{figure}[htp]
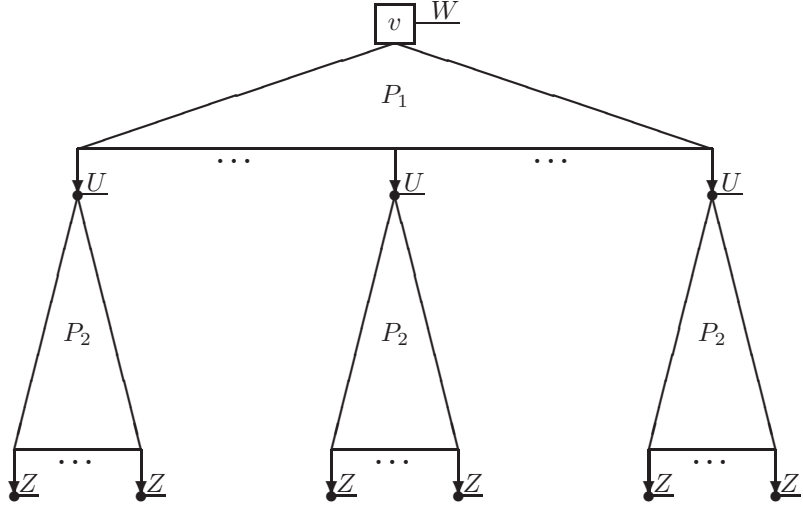

\begin{center}  \vPICTURE {\LOW} {3}{4}{6} {48}
\end{center}
\caption {{\bf Cut\/} as Composition of programs}   
\label{fcompos}
\end{figure}
 {\bf Case~of~Rule~Cut.}
 Suppose that $P_1$ and~$P_2$ are strong solutions to sequents
 of the form
     $$ \seq{W,\, \Gamma_1,\, \bang{\Delta_1}}{U} $$
 and $$ \seq{U,\, \Gamma_2,\, \bang{\Delta_2}}{Z},$$
 respectively.

 Now we can construct a Horn program~$P'$ with the help of the
 following operation of {\bf Composition}\ (see~Figure~\ref{fcompos}):
\begin{bbize}{a}
\item We glue each output vertex of~$P_1$ to the root
  of a copy of the program\/~$P_2$.
\end{bbize}
 It is clear that such a Horn program~$P'$ is a strong
 solution to the sequent
  $$ \seq{W,\,\Gamma_1,\,\Gamma_2,\,
            \bang{\Delta_1},\,\bang{\Delta_2}}{Z}.$$%

 {\bf The rest of the Cases.} Given a Horn program~$P_0$ that is a
 strong solution to a sequent representing the premise for one of
 the remaining rules, the same Horn program~$P_0$ can be considered
 as a strong solution to the corresponding conclusion sequent.
\QED


\section {FYI: Encoding Minsky Machines}

 The well-known non-deterministic \mbox{$n$-counter} machines
 are defined as follows.

 Minsky machines deal with $n$~counters that can
 contain non-negative integers.
 The current value of an \mbox{$m$-th counter} will be represented
 by the variable~$x_m$.
 This value
\begin{bbize}{a}
\item can be increased by 1,
      which is represented by the {\em assignment operation\/}
      \mbox{$ x_m:=x_m+1;$}
\item or can be decreased by 1,
      which is represented by the {\em assignment operation\/}
      \mbox{$ x_m:=x_m-1;$}
\end{bbize}

\begin{definition} \label{dM}
 The program of an \mbox{$n$-counter} machine~$M$ is a finite list
 of {\em instructions\/}
     $$ I_1;\ I_2;\ \ldots ;\ I_s; $$
 labelled by {\em labels}
     $$ L_0,\ L_1,\ L_2,\ \ldots ,\ L_i,\ \ldots ,\ L_j,\ \ldots $$
 Each of these instructions is of one of the following five types:\\
$\begin{array}{@{}cl}%
(1) & L_i:\ x_m:=x_m+1;\ \mbox{{\bf goto}}\ L_j;\\%
(2) & L_i:\ x_m:=x_m-1;\ \mbox{{\bf goto}}\ L_j;\\%
(3) & L_i:\ \mbox{{\bf if}}\ (x_m>0)\ \mbox{{\bf then goto}}\ L_j;\\%
(4) & L_i:\ \mbox{{\bf if}}\ (x_m=0)\ \mbox{{\bf then goto}}\ L_j;\\%
(5) & L_0:\ \mbox{{\bf halt;}}%
  \end{array}$\\
 where $L_i$ and $L_j$ are {\em labels,}  and \mbox{$i \geq 1.$}
\end{definition}

\begin{definition}
 An {\em instantaneous description (configuration)} is a tuple:
            $$ \mconf{L}{c_1}{c_2}{c_n} $$
 where $L$~is a label,\\
 \mbox{$ c_1,\ c_2,\ \ldots ,\ c_n $}\ \ are the current values of
 our $n$~counters, respectively.

 \noindent
 A {\em computation\/} of a Minsky machine~$M$
 is a (finite) sequence of  configurations
    $$ K_1,\ K_2,\ \ldots ,\ K_t,\ K_{t+1},\ \ldots , $$
 such that each {\em move} (from~$K_{t}$ to~$K_{t+1}$) can be performed
 by applying an instruction from the program of machine~$M$.
\end{definition}

\subsection {The Main Encoding}
 In our encoding we will use the following {\em literals:}
\begin{bbize}{i}
\item \mbox{$ r_1,\ r_2,\ \ldots ,\ r_m,\ \ldots ,\ r_n \ $}\footnote
      { Literal~$r_\good{m}$ is associated with the \mbox{$m$-th counter.}}
\item \mbox{$ l_0,\ l_1,\ l_2,\ \ldots ,\ l_i,\ \ldots ,\ l_j,\ \ldots $}%
      \footnote{ Literal~$l_\good{i}$ represents label~$L_\good{i}$.}
\item \mbox{$\kappa_1,\ \kappa_2,\ \ldots ,\ \kappa_m,\ \ldots ,\
                                                \kappa_n\ $}\footnote
      { Literal~$\kappa_\good{m}$ will be used to {\em kill\/} all
        literals except~$r_\good{m}$.}
\end{bbize}
 Each instruction~$I$ from the list of instructions~\mbox{(1)-(4)}
 of Definition~\ref{dM} will be axiomatized by the corresponding
 Linear Logic formula~$\varphi_\good{I}$ in the following way:\\
$\begin{array}{@{}lcl}%
 \varphi_{(1)} &=& \limply{l_i}{(l_j \otimes r_m)},\\%
 \varphi_{(2)} &=& \limply{(l_i \otimes r_m)}{l_j},\\%
 \varphi_{(3)} &=& \limply{(l_i \otimes r_m)}{(l_j \otimes r_m)},\\%
 \varphi_{(4)} &=& \lvariant{l_i}{l_j}{\kappa_m}.%
  \end{array}$\\[1ex]
 For a given machine~$M$, its program
      $$ I_1;\ I_2;\ \ldots ;\ I_s; $$
 is axiomatized by a multiset~$\Phi_M$ as follows:
 $$ \Phi_M \ = \ \varphi_\good{I_1},\  \varphi_\good{I_2},\ \ldots ,\
                 \varphi_\good{I_s}.$$
 In addition, for every~$m$, by~${\cal K}_\good{m}$ we mean the multiset
 consisting of one Horn implication:
          $$ \limply{\kappa_m}{l_0} $$
 and the following \ \mbox{$(n-1)$} \ \red{{\em killing\/}}
 implications:
 $$ \limply{(\kappa_m \otimes r_i)}{\kappa_m},\hspace{4em} (i \neq m) $$
 We set that
 $$ {\cal K} = \bigcup_\good{m=1}^\good{n} {\cal K}_\good{m} $$
 We will prove that an {\bf exact} correspondence
 exists between arbitrary
 computations of~$M$ on inputs
            $$  k_1,\ k_2,\ \ldots ,\ k_n $$
 and derivations for a sequent of the form
 $$ \seq{\lconf{l_1}{k_1}{k_2}{k_n},\,\bang{\Phi_M},\,\bang{{\cal K}}}{l_0}.$$
 More precisely, taking into account our complete computational
 interpretation for sequents of this kind
 \mbox{(Theorem~\ref{tlinprog}),}
 we will prove an {\bf exact} correspondence between arbitrary
 computations of~$M$ on inputs
            $$  k_1,\ k_2,\ \ldots ,\ k_n $$
 and {\em tree-like\/} strong solutions to this sequent
 $$ \seq{\lconf{l_1}{k_1}{k_2}{k_n},\,\bang{\Phi_M},\,\bang{{\cal K}}}{l_0}.$$
 In particular, each configuration~$K$
       $$ K = \mconf{L_i}{c_1}{c_2}{c_n} $$
 will be represented in Linear Logic by a simple tensor product
       $$ \widetilde{K} = \lconf{l_i}{c_1}{c_2}{c_n}.$$

\newcommand{ \COMP }[1]%
{\begin{picture}(0,0) \thicklines
\Ycur=21
\put(0,-\Ycur){\NodeXIV {$0$}
 {\mconf{L_1}{k_1}{k_2}{k_n}\ \ \ }{\toLeft}}  
\advance \Ycur by 14
\put(0,-\Ycur){\SouthD{#1}}
\advance \Ycur by #1
\put(0,-\Ycur){\NodeXIV {$u$}
 {\mconf{L'}{a_1}{a_2}{a_n}\ \ \ }{\toLeft}}
\advance \Ycur by 14
\put(0,-\Ycur){\SouthD{#1}}
\advance \Ycur by #1
\put(0,-\Ycur){\NodeXIV {$u'$}
 {\mconf{L''}{b_1}{b_2}{b_n}\ \ \ }{\toLeft}}
\advance \Ycur by 14
\put(0,-\Ycur){\SouthD{#1}}
\advance \Ycur by #1
\put(0,-\Ycur){\NodeXIV {}
 {\mconf{L'''}{c_1}{c_2}{c_n}\ \ \ }{\toLeft}}
\advance \Ycur by 14
\put(0,-\Ycur){\SouthD{#1}}
\advance \Ycur by #1
\put(0,-\Ycur){\NodeXIV {$t$}
 {\mconf{L_0}{0}{0}{0}\ \ \ }{\toLeft}}
\end{picture}}

\newcommand{ \PROG }[1]%
{\begin{picture}(0,0) \thicklines
\Ycur=21
\put(0,-\Ycur){\NodeXIV {$v_0$}
 {\lconf{l_1}{k_1}{k_2}{k_n}\ }{\toLeft}}  
\advance \Ycur by 14
\put(0,-\Ycur){\SouthD{#1}}
\advance \Ycur by #1
\put(0,-\Ycur){\NodeXIV {$v_u$}
 {\lconf{l'}{a_1}{a_2}{a_n}\ }{\toLeft}}
\advance \Ycur by 14
\put(0,-\Ycur){\SouthD{#1}}
\put(0,-\Ycur){\SouthEastD{#1}}
\advance \Ycur by #1
\put(0,-\Ycur){\NodeXIV {$v_{u'}$}
 {\lconf{l''}{b_1}{b_2}{b_n}\ }{\toLeft}}
\put(#1,-\Ycur){\NodeXIV {}{\ \ \ $l_0$}{\toRight}}
\advance \Ycur by 14
\put(0,-\Ycur){\SouthD{#1}}
\put(0,-\Ycur){\SouthEastD{#1}}
\advance \Ycur by #1
\put(0,-\Ycur){\NodeXIV {}
 {\lconf{l'''}{c_1}{c_2}{c_n}\ }{\toLeft}}
\put(#1,-\Ycur){\NodeXIV {}{\ \ \ $l_0$}{\toRight}}
\advance \Ycur by 14
\put(0,-\Ycur){\SouthD{#1}}
\put(0,-\Ycur){\SouthEastD{#1}}
\advance \Ycur by #1
\put(0,-\Ycur){\NodeXIV {$v_t$}
 {\lconf{l_0}{0}{0}{0}\ \ \ }{\toLeft}}
\put(#1,-\Ycur){\NodeXIV {}{\ \ \ $l_0$}{\toRight}}
\end{picture}}

\begin{figure*}[htp]
\begin{center}
 $\stackrel{\fbox{\vPICTURE {\COMP} {4}{8}{8} {40}}}{\mbox{(a)}}$\
 $\stackrel{\fbox{\vPICTURE {\PROG} {4}{8}{8} {40}}}{\mbox{(b)}}$
\end{center}
\caption {The correspondence:
          \mbox{Computation~(a)\ \ --- \ \ Horn Program~(b).}}
\label{fPROG}
\end{figure*}

\subsection {FYI: From computations to tree-like Horn programs}

\begin{lemma} \label{lMLL}
 For given inputs\ \mbox{$ k_1, k_2, \ldots, k_n$},
\ let~$M$ be able to go
 from the {\em initial\/} configuration
 \mconf{L_1}{k_1}{k_2}{k_n} to the
 {\em halting\/} configuration \mconf{L_0}{0}{0}{0}.

 Then we can construct a\/ {\em tree-like\/} Horn program~$P$,
 which is a strong solution to the sequent
 $$\seq{\lconf{l_1}{k_1}{k_2}{k_n},
 \,\bang{\Phi_M},\,\bang{{\cal K}}}{l_0}$$%
\comment{
 Then a sequent of the form
$$\seq{\lconf{l_1}{k_1}{k_2}{k_n},\,\bang{\Phi_M},
\,\bang{{\cal K}}}{l_0}$$%
 is derivable in Linear Logic.
}
\end{lemma}
\noindent{\bf Proof.}\  Let
  $$ K_0, \ K_1, \ K_2, \ \ldots,\ K_u,\ K_{u+1},\ \ldots,\ K_t $$
 be a computation of~$M$ \mbox{(See Figure~\ref{fPROG}(a))}
 leading from the
 initial configuration~$K_0$:
       $$ K_0 = \mconf{L_1}{k_1}{k_2}{k_n},$$
 to the halting configuration~$K_t$:
       $$ K_t = \mconf{L_0}{0}{0}{0}.$$
 Running from the beginning of this sequence of configurations
 to its end, we will construct a {\em tree-like\/} Horn program~$P$,
 which is a strong solution to the sequent
$$\seq{\lconf{l_1}{k_1}{k_2}{k_n},
 \,\bang{\Phi_M},\,\bang{{\cal K}}}{l_0},$$%
 and has the following peculiarities \mbox{(See Figure~\ref{fPROG}(b))}
\begin{bbize}{i}
\item $ P(\widetilde{K_0}) = \widetilde{K_t} = l_0,$
\item and, moreover, there exists a branch of~$P$,
 we call it {\em main\/}:
  $$ v_0, \ v_1, \ v_2, \ \ldots,\ v_u,\ v_{u+1},\ \ldots,\ v_t,$$
      such that for each vertex~$v_u$ from this {\em main\/} branch:
         $$\OUT{P,\widetilde{K_0},v_u} \cong \widetilde{K_u}.$$%
\end{bbize}
 We construct the desired program~$P$ by induction:\\
 The root~$v_0$ of~$P$ is associated with the
 initial configuration~$K_0$:
    $$\OUT{P,\widetilde{K_0},v_0} = \lconf{l_1}{k_1}{k_2}{k_n}.$$
 Let $v_u$ be the terminal vertex of the fragment of~$P$
 (that has already been constructed), associated with the current
 configuration~$K_u$:
      $$\OUT{P,\widetilde{K_0},v_u} \cong \widetilde{K_u}
               = \lconf{l_i}{a_1}{a_2}{a_n}.$$
 The move from~$K_u$ to~$K_{u+1}$ is simulated in the following way:
\begin{bbize}{a}
\item If this move is performed by applying an
 {\em assignment operation\/}
      instruction~$I$ from the list of instructions~\mbox{(1)-(3)} of
      Definition~\ref{dM},
      then we create a new edge~$(v_u,v_{u+1})$ and label this new edge
      by the corresponding Horn implication~$\varphi_\good{I}$, getting
      for this new terminal vertex~$v_{u+1}$:
      $$\OUT{P,\widetilde{K_0},v_{u+1}} \cong \widetilde{K_{u+1}}.$$
      Figure~\ref{fASSIGN} shows the case where
 this instruction~$I$ is of
      the form\ \mbox{$ L_i:\ x_1:=x_1-1;\ \mbox{{\bf goto}}\ L_j $.}
\item Let the foregoing move be performed
       by applying a {\em ZERO-test\/}
      instruction~$I$ of the form~(4)
  $$ L_i:\ \mbox{{\bf if}}\ (x_m=0)\ \mbox{{\bf then goto}}\ L_j.$$%
      The definability conditions of this move provide that
               $$ a_m = 0.$$
      We extend the fragment of~$P$ (that has already been constructed)
      as follows \mbox{(See Figure~\ref{fZERO}):}\\
      First, we create two new outgoing edges~$(v_u,v_{u+1})$
      and~$(v_u,w_u)$, and label these new edges by the
 Horn implications
        $$ \limply{l_i}{l_j}\ \ \mbox{and}\ \ \limply{l_i}{\kappa_m},$$
      respectively.
      It is readily seen that
     $$\begin{array}{lclcl}%
       \OUT{P,\widetilde{K_0},v_{u+1}}&\cong&%
         \lconfe{l_j}{a_1}{a_2}{a_m}{a_n}&=&\widetilde{K_{u+1}},\\%
       \OUT{P,\widetilde{K_0},w_u}&\cong&%
               \lconfe{\kappa_m}{a_1}{a_2}{a_m}{a_n}.&&%
       \end{array} $$
      Then, we create a chain of $t_u$~new edges
        $$ (w_u, w^u_1),\ (w^u_1,w^u_2),\ \ldots,\
                          (w^u_\good{t_u-1},w^u_\good{t_u})$$
      where
        $$ t_u = a_1 + a_2 + \cdots + a_{m-1} + a_{m+1} + \cdots + a_n,$$
      and label these new edges by such Horn implications
      from~${\cal K}_\good{m}$ as to {\em kill\/} all occurrences of
      literals
       $$ r_1,\ r_2,\ \ldots ,\ r_{m-1},\ r_{m+1},\ \ldots ,\ r_n,$$
      and ensure that
        $$\OUT{P,\widetilde{K_0},w^u_\good{t_u}} \cong
                            \lconfe{\kappa_m}{0}{0}{a_m}{0}.$$
      Finally, we create a new edge~$(w^u_\good{t_u},w^u_\good{t_u+1})$,
      and label this new edge by the Horn implication
          $$ \limply{\kappa_m}{l_0}.$$%
      Taking into account that\ \ \mbox{$a_m = 0$,}\ \ for the terminal
      vertex~$w^u_\good{t_u+1}$ of the foregoing chain, we have:
       $$\OUT{P,\widetilde{K_0},w^u_\good{t_u+1}} =
          \lconfe{l_0}{0}{0}{a_m}{0} = l_0.$$
\end{bbize}
 Hence, for all terminal vertices~$w$, i.e.~both for the terminal
 vertex~$v_t$ of the {\em main\/} branch
 and for the terminal vertices of all
 auxiliary chains, we obtain that
 $$\OUT{P,\widetilde{K_0},w} = l_0 = \widetilde{K_t}.$$
 Thus, our inductive process results
 in a {\em tree-like} Horn program~$P$
 that is a strong solution to the sequent
 $$ \seq{\lconf{l_1}{k_1}{k_2}{k_n},\,\bang{\Phi_M},\,\bang{{\cal K}}}
        {l_0} $$
\comment{
 To complete Lemma~\ref{lMLL}, we need only recall that, by
 Theorem~\ref{tlinprog}, the latter sequent is derivable
 in Linear Logic.
} 
\QED

\newcommand{ \ASSIGNa }[1]%
{\begin{picture}(0,0) \thicklines
\Ycur=21
\Xcur=#1 \multiply \Xcur by 2
\put(0,-\Ycur){\NodeXIV {$0$}
 {\mconf{L_1}{k_1}{k_2}{k_n}\ \ \ }{\toLeft}}  
\advance \Ycur by 14
\put(0,-\Ycur){\SouthD{#1}}
\advance \Ycur by #1
\put(0,-\Ycur){\NodeXIV {$u$}
 {\mconf{L_i}{a_1}{a_2}{a_n}\ \ \ }{\toLeft}}
\advance \Ycur by 14
\put(0,-\Ycur){\South{#1}
 {\ \ \ \ $L_i:\ x_1:=x_1-1;\ \mbox{{\bf goto}}\ L_j$}
 {\toRight}}
\advance \Ycur by #1
\put(0,-\Ycur){\NodeXIV {$u'$}
 {\mconf{L_j}{a_1 \mbox{-} 1}{a_2}{a_n}\ \ \ }{\toLeft}}
\advance \Ycur by 14
\put(0,-\Ycur){\SouthD{#1}}
\advance \Ycur by #1
\put(0,-\Ycur){\NodeXIV {$t$}
 {\mconf{L_0}{0}{0}{0}\ \ \ }{\toLeft}}
\end{picture}}
\newcommand{ \ASSIGNb }[1]%
{\begin{picture}(0,0) \thicklines
\Ycur=21
\Xcur=#1 \multiply \Xcur by 2
\put(0,-\Ycur){\NodeXIV {$v_0$}
 {\lconf{l_1}{k_1}{k_2}{k_n}\ \ \ }{\toLeft}}  
\advance \Ycur by 14
\put(0,-\Ycur){\SouthD{#1}}
\advance \Ycur by #1
\put(0,-\Ycur){\NodeXIV {$v_u$}
 {\lconf{l_i}{a_1}{a_2}{a_n}\ \ \ }{\toLeft}}
\advance \Ycur by 14
\put(0,-\Ycur){\South{#1}{\ \ \ \ \limply{(l_i\otimes r_1)}{l_j}}{\toRight}}
\advance \Ycur by #1
\put(0,-\Ycur){\NodeXIV {$v_{u'}$}
 {\lconf{l_j}{a_1 \mbox{-} 1}{a_2}{a_n}\ \ }{\toLeft}}
\advance \Ycur by 14
\put(0,-\Ycur){\SouthD{#1}}
\advance \Ycur by #1
\put(0,-\Ycur){\NodeXIV {$v_t$}
 {\lconf{l_0}{0}{0}{0}\ \ \ }{\toLeft}}
\end{picture}}
\begin{figure*}[htp]
\begin{center}
 $\stackrel{\fbox{\vPICTURE {\ASSIGNa} {3}{6}{7} {52}}}{\mbox{(a)}}$\
 $\stackrel{\fbox{\vPICTURE {\ASSIGNb} {3}{6}{7} {52}}}{\mbox{(b)}}$
\end{center}
\caption {The {\em assignment operation\/} correspondence:
          \mbox{(a)\ --\ (b).}}
\label{fASSIGN}
\end{figure*}

\newcommand{ \ZEROa }[1]%
{\begin{picture}(0,0) \thicklines
\Ycur=21
\Xcur=#1 \multiply \Xcur by 2
\put(0,-\Ycur){\NodeXIV {$0$}
 {\mconf{L_1}{k_1}{k_2}{k_n}\ \ \ }{\toLeft}}  
\advance \Ycur by 14
\put(0,-\Ycur){\SouthD{#1}}
\advance \Ycur by #1
\put(0,-\Ycur){\NodeXIV {$u$}
 {\mconf{L_i}{a_1}{a_2}{a_n}\ \ \ }{\toLeft}}
\advance \Ycur by 14
\put(0,-\Ycur){\South{#1}
 {\ \ \ \ $L_i:\mbox{{\bf if}}\ (x_m=0)\ \mbox{{\bf then goto}}\ L_j$}
 {\toRight}}
\advance \Ycur by #1
\put(0,-\Ycur){\NodeXIV {$u'$}
 {\mconf{L_j}{a_1}{a_2}{a_n}\ \ \ }{\toLeft}}
\advance \Ycur by 14
\put(0,-\Ycur){\SouthD{#1}}
\advance \Ycur by #1
\put(0,-\Ycur){\NodeXIV {$t$}
 {\mconf{L_0}{0}{0}{0}\ \ \ }{\toLeft}}
\end{picture}}
\newcommand{\ZEROb}[1]%
{\begin{picture}(0,0) \thicklines
\Ycur=21
\Xcur=#1 \multiply \Xcur by 2
\put(0,-\Ycur){\NodeXIV {$v_0$}
 {\lconf{l_1}{k_1}{k_2}{k_n}\ \ \ }{\toLeft}}  
\advance \Ycur by 14
\put(0,-\Ycur){\SouthD{#1}}
\advance \Ycur by #1
\put(0,-\Ycur){\NodeXIV {$v_u$}
 {\lconf{l_i}{a_1}{a_2}{a_n}\ \ \ }{\toLeft}}
\advance \Ycur by 14
\put(0,-\Ycur){\South    {#1}{\limply{l_i}{l_j}\ }{\toLeft}}
\put(0,-\Ycur){\SouthEast{#1}{\ \ \ \limply{l_i}{\kappa_m}}{\toRight}}
\advance \Ycur by #1
\put(0,-\Ycur){\NodeXIV {$v_{u'}$}
 {\lconf{l_j}{a_1}{a_2}{a_n}\ \ \ }{\toLeft}}
\put(#1,-\Ycur){\NodeXIV {$w_u$}
 { \lconf{\kappa_m}{a_1}{a_2}{a_n}}{\toRight}}
\advance \Ycur by 14
\put(0,-\Ycur){\SouthD{#1}}
\put(#1,-\Ycur){\SouthEastD{#1}}
\advance \Ycur by #1
\put(0,-\Ycur){\NodeXIV {$v_t$}
 {\lconf{l_0}{0}{0}{0}\ \ \ }{\toLeft}}
\put(\Xcur,-\Ycur){\NodeXIV {}{\ \ \ $l_0$}{\toRight}}
\end{picture}}
\begin{figure*}[htp]
\begin{center}
 $\stackrel{\fbox{\vPICTURE {\ZEROa} {3}{6}{8} {52}}}{\mbox{(a)}}$\
 $\stackrel{\fbox{\vPICTURE {\ZEROb} {3}{6}{8} {52}}}{\mbox{(b)}}$
\end{center}
\caption {The {\em ZERO-test\/} correspondence: \mbox{(a)\ --\ (b).}}
\label{fZERO}
\end{figure*}

\section{From tree-like Horn programs to Minsky computations}
\label{s-programs-Minsky} 

{
\begin{theorem}\label{t-lLLM}
 For given integers\ \mbox{$ k_1, k_2, \ldots, k_n,$}
 let a sequent of the form
$$\seq{\lconf{l_1}{k_1}{k_2}{k_n},\,\bang{\Phi_M},\,
 \bang{{\cal K}}}{l_0}$$%
 be derivable in Linear Logic. Then~$M$ can go from an
 {\em initial\/} configuration of the form
   $$\mconf{L_1}{k_1}{k_2}{k_n}$$%
 to
 the {\em halting\/} configuration
 $$ \mconf{L_0}{0}{0}{0}.$$%
\end{theorem}
\noindent{\bf Proof.}
 By Theorem~\ref{t-ll-HLL} and Theorem~\ref{tlinprog},
 we can construct a {\em tree-like} Horn program~$P$ such that
\begin{bbize}{i}
\item Each of the Horn implications occurring in~$P$ is drawn either
      from~$\Phi_M$ or from~${\cal K}$.
\item \rede{For all terminal vertices~$w$ of~$P$:
            $$ \OUT{P,W_0,w} = l_0 $$}%
      where
        $$ W_0 = \lconf{l_1}{k_1}{k_2}{k_n}.$$
\end{bbize}
\begin{lemma}\label{lLLM1}
 Running from the root~$v_0$ to terminal vertices of~$P$,
 we assemble the desired Minsky computation as follows:
\begin{bbize}{a}
\item It is proved that program~$P$ cannot be but of the form represented
      in Figure~\ref{fPROG}.
\item We can identify a branch of~$P$, called the {\em main\/} branch:
      $$ v_0, \ v_1, \ v_2, \ \ldots,\ v_u,\ v_{u+1},\ \ldots,\ v_t,$$
      such that for all vertices~$v_u$ on this branch,
       \OUT{P,W_0,v_u} is proved to be of the form
       $$ \OUT{P,W_0,v_u} \cong \lconf{l_i}{a_1}{a_2}{a_n}.$$
\item For all non-terminal vertices~$w'$ of~$P$ that are outside the
      {\em main\/} branch,
       \OUT{P,W_0,w'} is proved to be of the form
       $$ \OUT{P,W_0,w'} \cong \lconf{\kappa_m}{a_1}{a_2}{a_n}.$$
\item Finally, the following sequence of configurations
      \mbox{(See Figure~\ref{fPROG})}
      $$ K_0, \ K_1, \ K_2, \ \ldots,\ K_u,\ K_{u+1},\ \ldots,\ K_t $$
      such that for every integer~$u$
       $$ \widetilde{K_u} \cong \OUT{P,W_0,v_u},$$
      is proved to be a {\em successful\/} computation of~$M$ leading from
      the initial configuration~$K_0$:
       $$ K_0 = \mconf{L_1}{k_1}{k_2}{k_n},$$
      to the halting configuration~$K_t$:
       $$ K_t = \mconf{L_0}{0}{0}{0}.$$
\end{bbize}
\end{lemma}

\proof\ Since
   $$ \widetilde{K_0} = \OUT{P,W_0,v_0} = \lconf{l_1}{k_1}{k_2}{k_n},$$
 we have for the root~$v_0$:
       $$ K_0 = \mconf{L_1}{k_1}{k_2}{k_n}.$$
 Let $v_u$ be the current vertex on the {\em main\/} branch we are searching
 for, and, according to the inductive hypothesis, let \OUT{P,W_0,v_u}
 be of the form
 $$ \OUT{P,W_0,v_u} \cong \widetilde{K_u} = \lconf{l_i}{a_1}{a_2}{a_n}.$$
 There are the following cases to be considered.
\begin{bbize}{a}
\item Suppose that $v_u$ is a non-divergent vertex with the only
      son which will be named~$v_{u+1}$.
      According to the definability conditions for our program~$P$,
      the single outgoing edge~$(v_u,v_{u+1})$ cannot be labelled but
      by a Horn implication~$A$ from~$\Phi_M$.
      Moreover,
               $$ A = \varphi_\good{I} $$
      for some instruction~$I$ of the form~\mbox{(1)-(3)} from
      Definition~\ref{dM}. Let this instruction~$I$ be of the form
        $$ L_i:\ x_1:=x_1-1;\ \mbox{{\bf goto}}\ L_j;$$
      and
        $$ A = \limply{(l_i \otimes r_1)}{l_j}.$$
      Then we have \mbox{(See Figure~\ref{fASSIGN}):}
   $$ \OUT{P,W_0,v_{u+1}} \cong \lconf{l_j}{a_1 \mbox{-} 1}{a_2}{a_n},$$
      (and, hence,\ \ \mbox{$a_1 \geq 1$).}\ \
      Performing the foregoing instruction~$I$, machine~$M$ can move from
      the current configuration~$K_u$:
        $$ K_u = \mconf{L_i}{a_1}{a_2}{a_n},$$
      to the next configuration~$K_{u+1}$:
        $$ K_{u+1} = \mconf{L_j}{a_1 \mbox{-} 1}{a_2}{a_n},$$
      such that
       $$ \widetilde{K_{u+1}} \cong \OUT{P,W_0,v_{u+1}}.$$
      The remaining cases are handled similarly.
\item The {\bf crucial point} is where $v_u$ is a vertex with two outgoing
      edges, say~$(v_u,v_{u+1})$ and~$(v_u,w_u)$, labelled by Horn
      implications~$A_1$ and~$A_2$, respectively.
      \mbox{(See Figure~\ref{fZERO})}

      It means that the implication used at this point of program~$P$
      must be a \pH implication~$A$ from~$\Phi_M$ of the form
        $$ A = \lvariant{l_i}{l_j}{\kappa_m},$$
      and, in addition,
      $$\begin{array}{lcl}%
         A_1 & = & \limply{l_i}{l_j} \\%
         A_2 & = & \limply{l_i}{\kappa_m}.
        \end{array} $$
      Therefore,
    $$\begin{array}{lcl}%
       \OUT{P,W_0,v_{u+1}} & \cong & \lconfe{l_j}{a_1}{a_2}{a_m}{a_n}\\%
       \OUT{P,W_0,w_u} & \cong & \lconfe{\kappa_m}{a_1}{a_2}{a_m}{a_n}.
      \end{array} $$
     Let us examine the descendants of the vertex~$w_u$.

     Taking into account the definability conditions,
     any edge~$(w_1,w_2)$, such that $w_1$~is a descendant of~$w_u$,
     cannot be labelled but by a Horn implication from~${\cal K}_m$.

     So we can conclude that for all non-terminal descendants~$w'$ of the
     vertex~$w_u$,\ \OUT{P,W_0,w'} is of the form
       $$ \OUT{P,W_0,w'} \cong \lconfe{\kappa_m}{c_1}{c_2}{a_m}{c_n}.$$
     For the terminal descendant~$w$ of the vertex~$w_u$, \OUT{P,W_0,w}
     is to be of the form
       $$ \OUT{P,W_0,w} \cong \lconfe{l_0}{c_1}{c_2}{a_m}{c_n}.$$
     Recalling that
               $$ \OUT{P,W_0,w} = l_0,$$
     we get the desired
               $$ a_m = 0.$$
     Indeed,
            $$ A = \varphi_\good{I} $$
     for a {\em ZERO-test} instruction~$I$ of the form
      $$ L_i:\ \mbox{{\bf if}}\ (x_m=0)\ \mbox{{\bf then goto}}\ L_j;$$
     Performing this instruction~$I$, machine~$M$ can move from the
     current configuration~$K_u$:
        $$ K_u = \mconfe{L_i}{a_1}{a_2}{a_m}{a_n},$$
     to the next configuration~$K_{u+1}$:
        $$ K_{u+1} = \mconfe{L_j}{a_1}{a_2}{a_m}{a_n},$$
     such that
       $$ \widetilde{K_{u+1}} \cong \OUT{P,W_0,v_{u+1}}.$$
\item Suppose that the {\em main\/} branch we have been developing
      $$ v_0, \ v_1, \ v_2, \ \ldots,\ v_u,\ v_{u+1},\ \ldots,\ v_t,$$
      has ended at a vertex~$v_t$. According to what has been said,
       $$ l_0 = \OUT{P,W_0,v_t} \cong \widetilde{K_t}
                              = \lconf{l_j}{c_1}{c_2}{c_n}.$$
      Hence, this configuration~$K_t$ is the {\em halting\/} one:
          $$ K_t = \mconf{L_0}{0}{0}{0}.$$
\end{bbize}
 Now, bringing together all the cases considered, we can complete
 Lemma~\ref{lLLM1} and, hence, Theorem~\ref{t-lLLM}.
\QED

\begin{theorem} \label{tMLL}
 For given inputs\ \mbox{$ k_1, k_2, \ldots, k_n,$}\ \
 \mbox{an $n$-counter} Minsky machine~$M$ can go from the {\em initial\/}
 configuration \mconf{L_1}{k_1}{k_2}{k_n} to the {\em halting\/}
 configuration \mconf{L_0}{0}{0}{0} {\bf if and only if} a sequent of
 the form
$$\seq{\lconf{l_1}{k_1}{k_2}{k_n},\,\bang{\Phi_M},\,\bang{{\cal K}}}{l_0}$$
 is derivable in Linear Logic.
\end{theorem}

\proof\ We bring together Lemma~\ref{lMLL} and Theorem~\ref{t-lLLM}.
\QED
} 





\end{document}